\documentclass{aastex62}

\usepackage{tablefootnote}
\usepackage{color, colortbl}
\definecolor{Gray}{gray}{0.9}
\definecolor{LightCyan}{rgb}{0.88,1,1}
\usepackage{xcolor}
\received{October 9, 2020}
\accepted{October 9, 2020}
\submitjournal{Philisophical Transactions A (Royal Society Publishing)}

\begin{document}
\title{Oscillations observed in Umbra, Plage, Quiet-Sun and the Polarity Inversion Line of Active Region 11158 using HMI/SDO Data}
\author{A. A. Norton}
\affiliation{Stanford University, Stanford, CA, USA 94305}
\author{R. B. Stutz}
\affiliation{Stanford University, Stanford, CA, USA 94305}
\author{B. T. Welsch}
\affiliation{University of Wisconsin, Green Bay, WI, USA 54311}
\correspondingauthor{A. A. Norton}
\email{AANorton@sStanford.EDU}


\begin{abstract}
Using data from the Helioseismic Magnetic Imager, we report on the amplitudes and phase relations of oscillations in quiet-Sun, plage, umbra and the polarity inversion line (PIL) of an active region NOAA$\#$11158. We employ Fourier, wavelet and cross correlation spectra analysis. Waves with 5--minute periods are observed in umbra, PIL and plage with common phase values of ${\phi}(v,I)=\frac{\pi}{2}$,
${\phi}(v,B_{los})=-\frac{\pi}{2}$. In addition, ${\phi}(I,B_{los})=\pi$ in plage are observed. These phase values are consistent with slow standing or fast standing surface sausage wave modes. The line width variations, and their phase relations with intensity and magnetic oscillations, show different values within the plage and PIL regions, which may offer a way to further differentiate wave mode mechanics.  Significant Doppler velocity oscillations are present along the PIL, meaning that plasma motion is perpendicular to the magnetic field lines, a signature of Alv\`enic waves. A time-distance diagram along a section of the PIL shows Eastward propagating Doppler oscillations converting into magnetic oscillations; the propagation speeds range between 2$-$6 km s$^{-1}$. Lastly, a 3-minute wave is observed in select regions of the umbra in the magnetogram data. 
\end{abstract}

\section{Introduction}

The Sun is a seething mass of plasma with a great variety of magnetic fields and electric currents being dynamically generated and distributed throughout its layers. MHD waves, generated either by mode conversion of $p$-modes or excited due to shaking of magnetic flux tubes by turbulent convection and other motions, are thought to contribute to shock heating in the chromospheric layers and, as such, are an important ingredient in heating of the upper atmosphere. Alfv\'en waves \citep{alfven:1947} are the least impeded of the MHD waves since they are not reflected by pressure gradients and therefore may reach the corona before dissipating \citep{mathioudakis:2013} and may play a role in the acceleration of the solar wind. 

 It is certain that magnetic flux tubes in the solar atmosphere host MHD waves because theory indicates there is no such thing as a pure acoustic wave in a magnetized plasma \citep{cally:2005} and also because local helioseismology shows that acoustic shadows exist downstream from sunspots \citep{braun:1987, braun:2008}, meaning that strong magnetic fields decrease the amount of acoustic wave power outgoing horizontally from the region of strong field compared to the acoustic wave power that was observed as incoming.  This decrease in acoustic wave power is explained by conversion of acoustic waves into upward or downward propagating MHD waves.    

In addition to observing a decrease in acoustic wave power exiting strong magnetic structures, local helioseismology techniques record time-travel changes in waves as they pass through magnetic regions. The changes in the travel times and phase shifts are caused by variations of the sound speed, sub-surface flows and magnetic properties of the medium through which the waves propagate, although it is difficult to differentiate the changes  caused by each property  \citep{werne:2004}. Mode conversion of acoustic waves into MHD waves can introduce phase shifts, too. Numerical modeling indicates that fast magnetic waves are created after acoustic waves are converted at the ${\beta}=1$ layer \citep{schunker:2006}. The fast magnetic waves are either reflected downward due to the gradient in the Alfvén speed \citep{khomenko:2006} or converted into upward and downward traveling Alfvén waves \citep{felipe:2012, khomenko:2012}.

MHD waves contain oscillations in the  magnetic field magnitude or direction, so one way to confirm their presence is to observe a time varying component of either. This is not the only way to confirm an MHD wave, though, since area oscillations \citep{grant:2015} and spectral line width oscillationsv\citep{jess:2009} have been used in tandem with intensity and velocity oscillations to confirm the presence of MHD waves. Confident detection of magnetic field oscillations is an observational challenge with reports of amplitudes being low, i.e., an upper amplitude of 4 Gauss as observed with the Advanced Stokes Polarimeter in sunspot field strengths for 5 minute oscillations \citep{lites:1998}, 6 Gauss in umbral regions \cite{Ruedi:1998}, 7$-$11 Gauss in umbra using an infrared line\cite{rubio:2000} with only a portion of this amplitude being due to magnetoacoustic waves \citep{khomenko:2003}, 4$-$17 Gauss in pores and network \citep{fujimura:2009}, and 20 Gauss in magnetic flux oscillations as estimated by  \citet{kanoh:2016} in a study using Hinode SP observations, among other reports in the literature that confirm small amplitudes for $\delta$B.  Such small wave amplitudes do not imply that the waves carry only small amounts of energy as the energy flux of an MHD wave is proportional to the original field strength, B, times the oscillating components, $\delta$B and $\delta$v, as described by the Poynting vector. 

In addition to searching for oscillations in the magnetic field, phase relations between different quantities are often used to interpret the oscillations as specific wave modes. This was outlined and observed by \citet{ulrich:1996} for magnetic flux and Doppler velocity at different positions on the solar disk and further studied by \citet{norton:2001,norton:2000}.  The observed phases may also indicate that instrumental cross talk \citep{settele:2002} or opacity fluctuations sampling the vertical magnetic gradient \citep{rubio:2000,norton:2002} are responsible for the periodic signal instead of MHD waves, although it should be noted that opacity fluctuations are themselves a signature of waves. A more recent development of expected phases for photospheric mode identification was provided for $B_{los}$, Doppler velocity and intensity for observations at a single atmospheric height by \citet{moreels:2013}.  Oscillations observed in pores, their area and intensity, with resulting phase relations, show upward propagating sausage modes as surface modes \citep{grant:2015}, and this work was furthered by \citet{keys:2018}, who concludes that surface waves are more prevalent than body waves in pores.  Multi-height studies by \citet{stangalini:2018} used circular polarization and intensity oscillations at the umbra-penumbra boundary to detect surface modes.  \citet{kanoh:2016} found evidence of standing slow-mode waves in the umbra with Hinode SP data based on phase analysis. Phase studies, both at single heights between different observables and at multiple heights, have been successful in identifying waves modes and characteristics. 

\begin{figure}[!t]
\centering
    \includegraphics[width=6.0in]{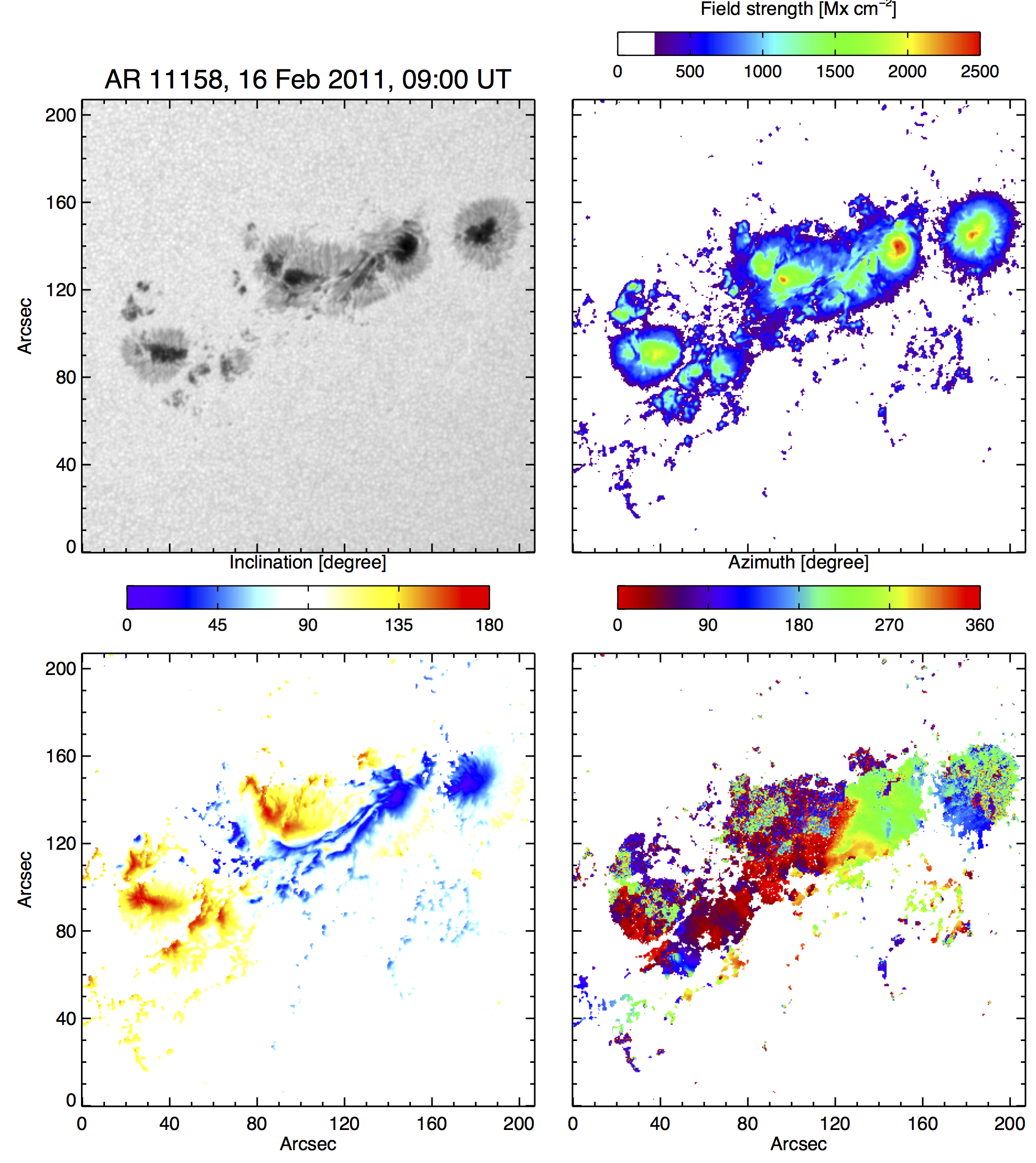}
  \caption{NOAA AR 11158 (HMI  Active  Region  Patch  377) is shown for 9:00 UT on 16 February 2011. The panels are the continuum intensity (upper left), total field strength (upper right), the inclination angle relative to the line of sight (lower left) with  field  directed  toward  the  observer  shown  in  blue,  transverse  field  in  white,  and  field  directed  away  from  the observer in red, and azimuth angle (lower right) adjusted to give the angle relative to the direction of rotation (i.e.west). Only pixels above a 250 Mx cm$^{-2}$ threshold are plotted.}
  \label{fig:vec}
\end{figure}

As the solar atmosphere is highly stratified, many waves are not capable of travelling into the outer atmosphere as they are reflected by density gradients, but magnetic fields and radiative losses allow more wave power to travel than originally thought. In general, waves in the quiet-Sun with frequencies less than the cutoff frequency of 5.2 mHz are considered trapped in or below the photosphere as resonant or standing waves.  However, in the presence of magnetic fields, the acoustic cut-off frequency in the photosphere is reduced due to temperature and density differences compared to the quiet-Sun. The cutoff frequency can be further reduced either by the inclination of the magnetic field with respect to the solar surface, by a factor of cos$\theta$ where $\theta$ is the inclination \citep{bel:1977}, or by radiative losses in thin vertical flux tubes such as faculae \citep{centeno:2009,khomenko:2008}. Cutoff frequencies in umbra and pores are on the order of $\sim$4 mHz and those in smaller magnetic structures that suffer significant radiative losses can be as low as 2 mHz, although recent numerical simulations \citep{felipe:2020} show that the cutoff frequencies in thin structures do not get quite that low. The exact reduction of cutoff frequencies due to radiative losses in magnetic structures is still an open question. The vertical gradients in the transition region offer another hindrance to wave propagation further into the corona. However, observed photospheric power in frequencies higher than the cutoff frequency can be considered part of the wave flux budget available to the chromosphere.

Importantly, the work by \citet{centeno:2006,centeno:2009}, who studied different types of solar structures (large and small sunspots, a pore and faculae region) at photospheric and chromospheric heights, showed that "while  the  atmospheric  cutoff  frequency and  the  propagation  properties  of  different  oscillating  modes  depend  on  the  magnetic  feature,  in  all  the cases  the  power  that  reaches  the  high  chromosphere  above  the  atmospheric  cutoff  comes  directly  from  the photosphere  by  means  of  linear  vertical  wave  propagation  rather  than  from  nonlinear  interaction  of  modes."  This indicates that the photospheric power present above the cutoff frequency does propagate into the chromosphere via linear vertical waves. 

\begin{figure}[!t]
\centering
\includegraphics[width=3.0in]{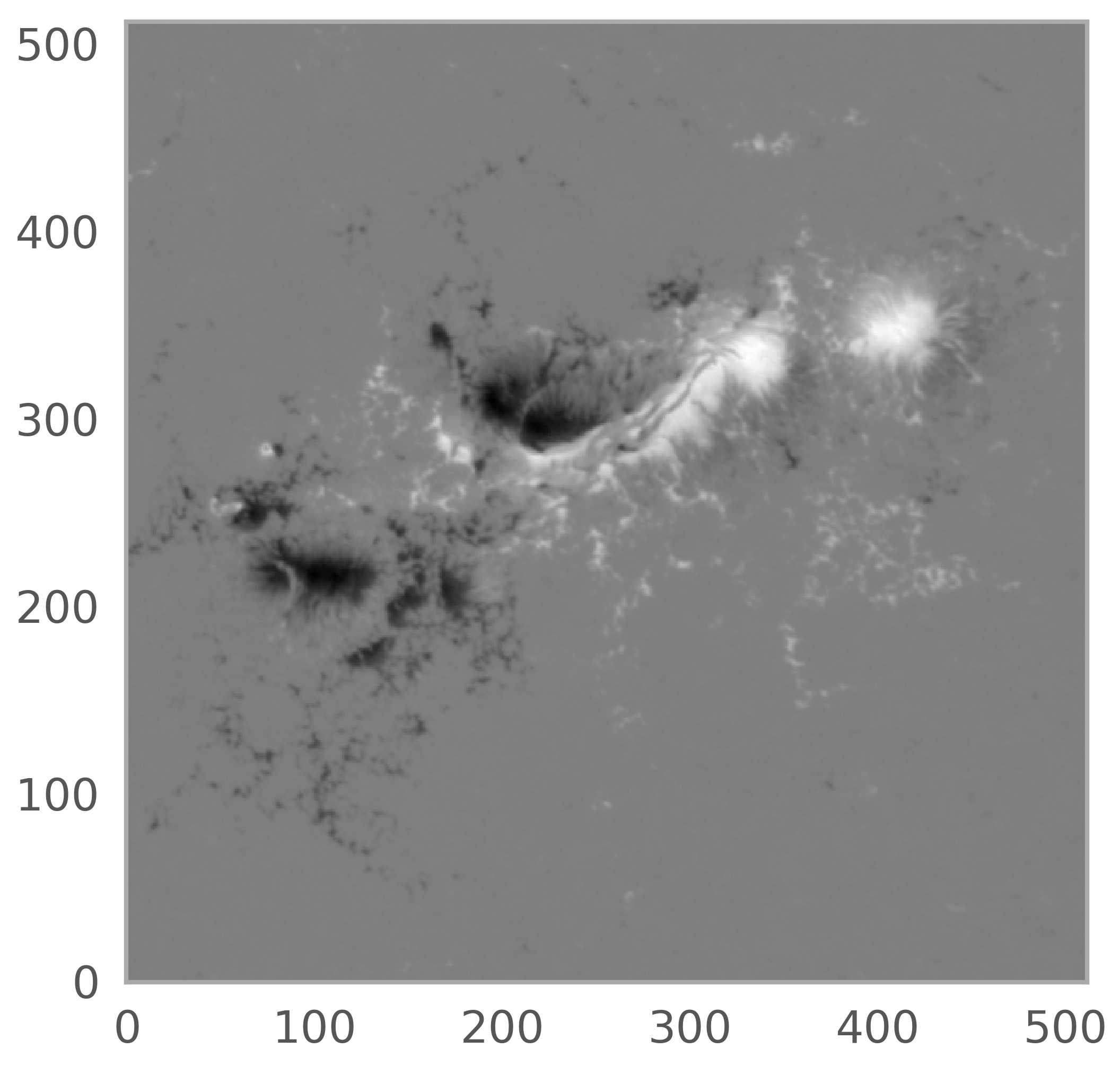}
\includegraphics[width=3.0in]{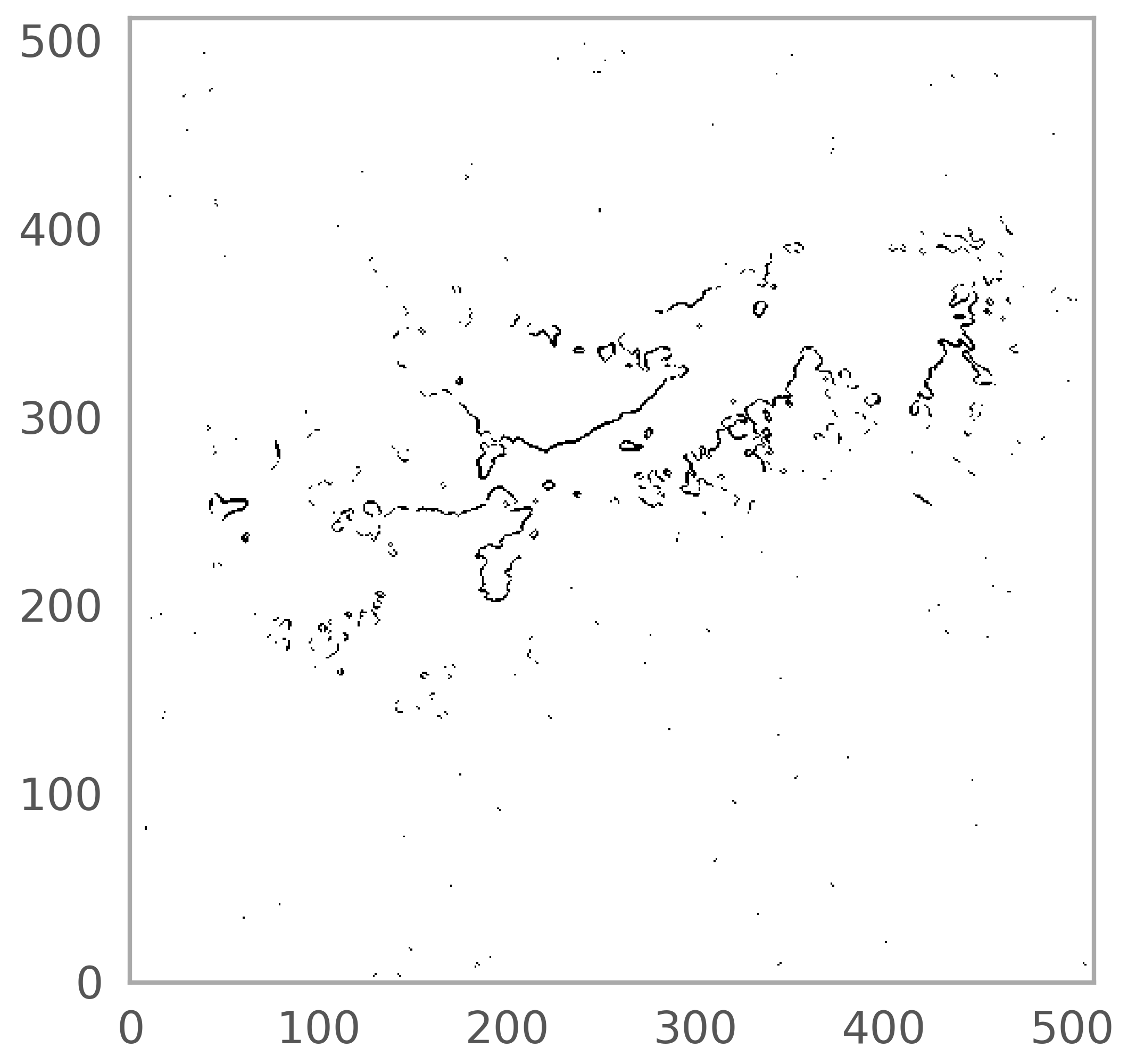}
\caption{Pixels within the AR 11158 region located along the polarity inversion line, or the neutral line, are shown (right panel) as determined using the line-of-sight magnetogram (at left) to identify which pixels are found with close proximity to both a positive and negative magnetogram value. The neutral line from $x=200-300$ and $y=280-320$ was stable for days, including the time period of our analysis, and did not experience an eruption.}
\label{fig:pil}
\end{figure}

While HMI data is often used for placing high-resolution observations in the context of a greater field-of-view, surprisingly few researchers have used HMI data for MHD wave studies.  Time-distance analysis of HMI data by \citet{cho:2020} determined the source depth of perturbations responsible for the slow mode waves seen in umbra to be $\sim$1000-2000 km below the photosphere. The detection of a fast-moving wave propagating from sunspot umbra through penumbra to about 15 Mm beyond the sunspot boundary by \citet{zhao:2015} was thought to be a magnetoacoustic wave, excited at $\sim$5 Mm beneath the sunspot’s surface. An analysis of waves in and above a sunspot with characteristics of p-modes using HMI and AIA \citet{zhao:2016} showed they were able to trace the waves upwards, and that the waves were able to channel through the chromosphere, transition region, and corona. The most extensive work on MHD waves using HMI data, in coordination with AIA data, has been carried out by \citet{rajaguru:2019, rajaguru:2013}, with the important findings that small scale magnetic elements channel significantly more acoustic wave energy into the chromosphere than originally expected.  These waves have frequencies between 2-5 mHz and the findings support \citet{centeno:2009,khomenko:2008,felipe:2020} in the argument that radiative losses, not magnetic inclinations, are responsible for lowering the cutoff frequencies for the thin magnetic structures and enable a large amount of wave flux to move upwards.
 
\begin{figure}[!ht]
\centering
\includegraphics[width=4.0in]{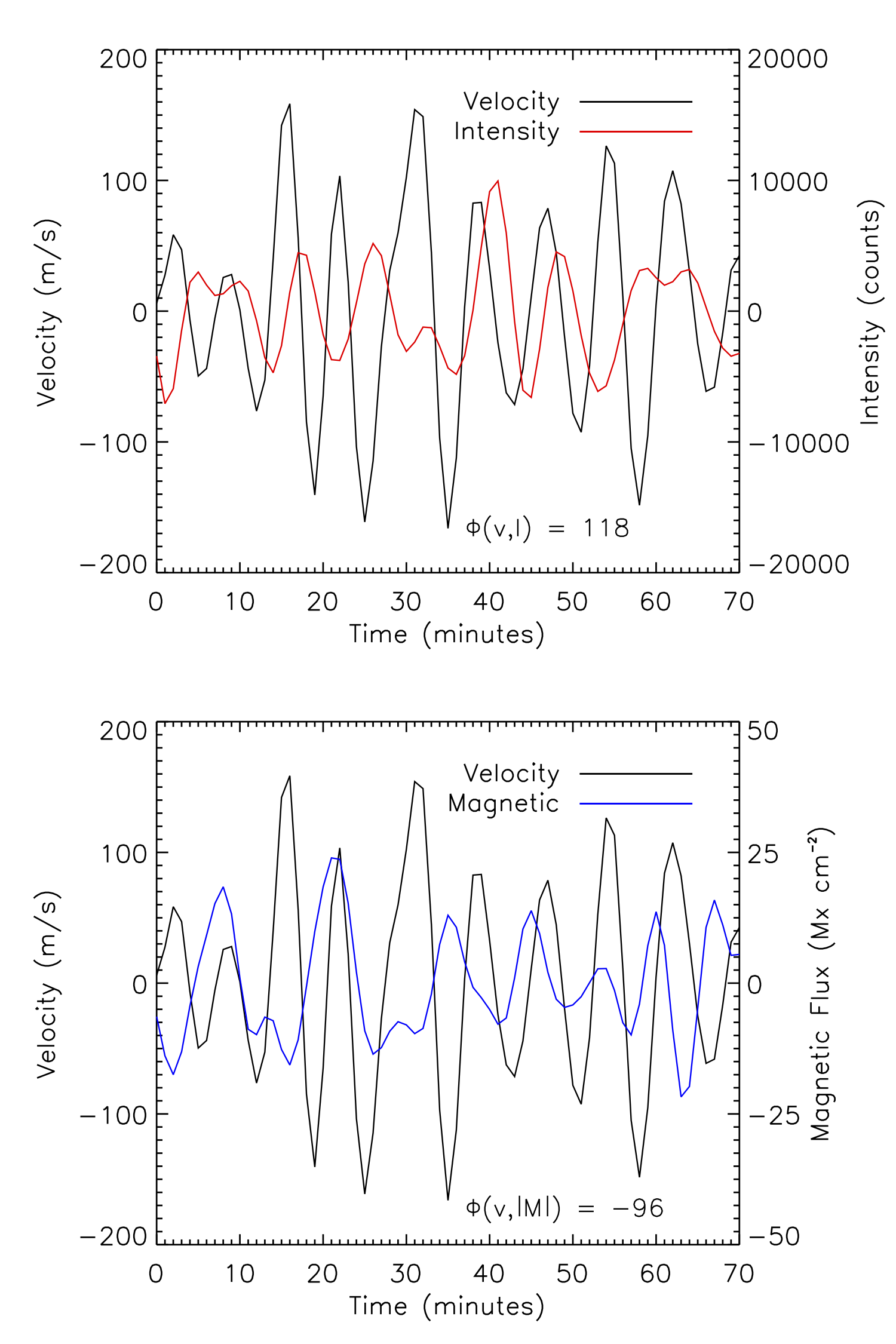}
\caption{Sample oscillations in the 2-4 mHz range for AR11158 for pixel location (230,288) in the polarity inversion line shows the velocity and intensity (top) and velocity and magnetic flux (bottom). As for all signal analysis in this paper, a 30-minute running average value was subtracted and the absolute value of the magnetic field was taken. The phase value of $\phi$(v,$\delta$$|$M$|$)=-96$^{\circ}$ indicates the velocity signal lags the magnetic signal by nearly $\frac{\pi}{2}$ whereas the $\phi$(v,$\delta$I)=118$^{\circ}$ value indicates that the velocity signal leads the intensity signal by a bit more than  $\frac{\pi}{2}$. }
\label{fig:osc}
\end{figure}

A search for Alfvénic waves using HMI data was suggested by Welsch (personal communication, 2017) to measure the Doppler velocities of the plasma in the vicinity of field lines that are perpendicular to the observer's line of sight, i.e. polarity inversion lines or neutral lines. Detecting oscillations transverse to magnetic field lines indicates the waves have properties of Alfv{\'e}nic waves.  HMI is designed with a high sensitivity to Doppler velocities whereas the line of sight magnetic field measurements of HMI are most likely too noisy to detect the photospheric amplitude of magnetic field changes associated with single mode MHD waves.  We should note that the nominal vector magnetic field data from HMI has a cadence of 720 seconds that precludes detection of the waves in the photosphere that have periods of five minutes or less, but the 135-second vector magnetic field data from HMI has a Nyquist frequency of 3.7 mHz so it can be used for wave research, but we discuss such data options further in Section 2. 

\begin{figure}[!t]
\centering
\includegraphics[width=5.5in]{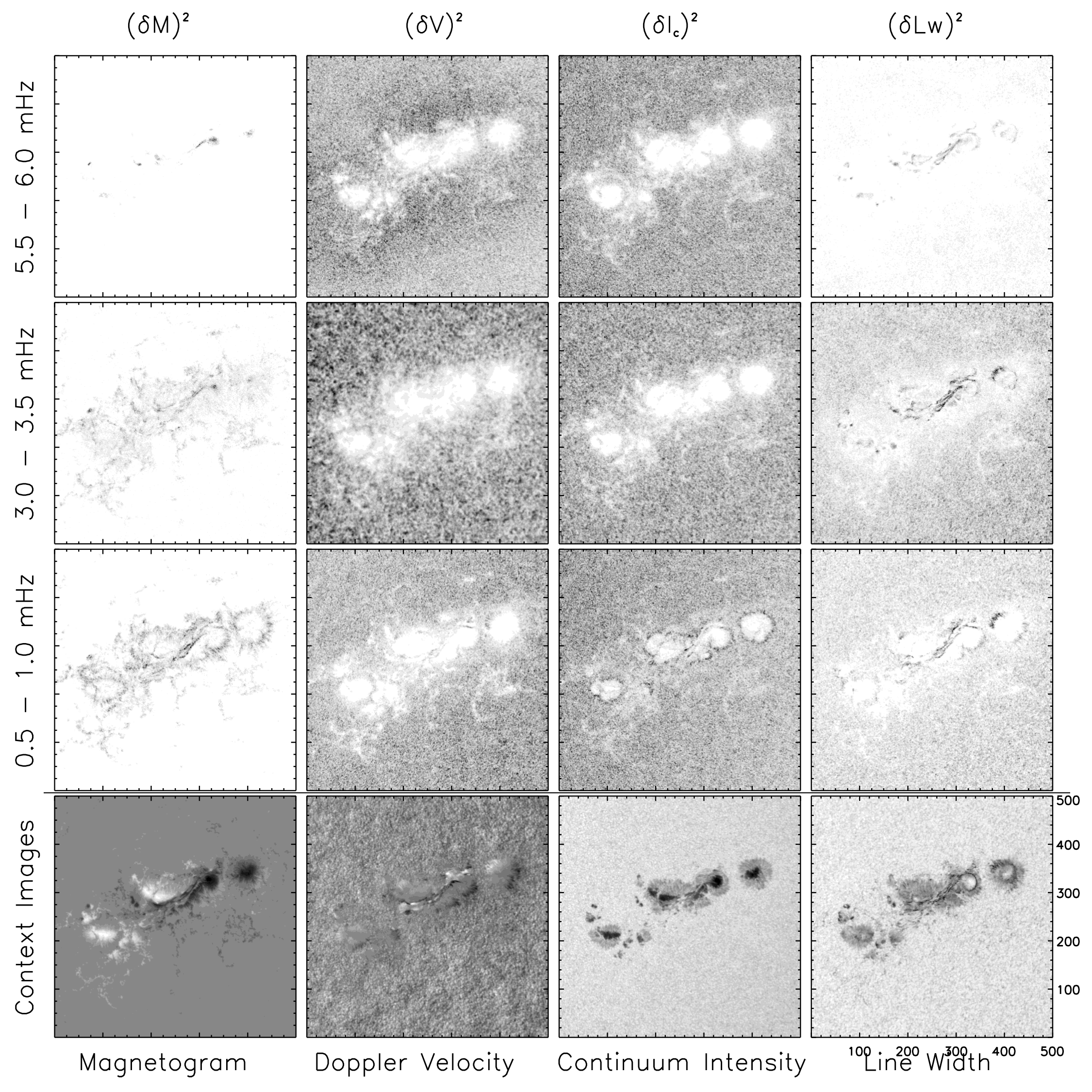}
\caption{Spatial distribution of power for AR 11158 as determined using classic Fourier analysis for an 8 hour time series. Columns from left to right are B$_{los}$, Doppler velocity, continuum intensity and line width.  Rows from top to bottom are the frequency ranges of 5.5$-$6.0, 3.0$-$3.5 and 0.5$-$1.0 mHz with the lowest row being a context image of the data. The frequencies represent the 3 minute (top row), 5 minute and 20 minute band which shows the dynamic evolution of the region. Black represents larger values excepting in the $I_c$ context image where black represents lower intensity. The context images from left to right correspond to parameter ranges of $-$2.0 $\le$ B$_{los}$ $\ge$ 2.3 kG, $-$2.3 $\le$ $v$ $\ge$ 2.3 km s$^{-1}$, 8000 $\le$ $I_{c}$ $\ge$ 60000 counts and 85 $\le$ Lw $\ge$ 201 m\AA. To aid in visual clarity, the Fourier power maps for each variable (left to right) are saturated at the following levels: B$_{los}$ $\le$ 2500, 400, and 285 G$^{2}$ Hz$^{-1}$ for the lowest to highest frequencies, v $\le$ 1$\times$10$^5$ m$^{2}$ s$^{-2}$ Hz$^{-1}$ for all frequencies, $I_c$ $\le$ 5$\times$10$^5$ counts$^{2}$ Hz$^{-1}$ for all frequencies, and Lw $\le$ 132, 19, and 11 m\AA$^{2}$ Hz$^{-1}$ for lowest to highest frequencies shown. }
\label{fig:spatial_pow}
\end{figure}

Recently, Alfv\'en waves heating chromospheric plasma in a sunspot umbra through the formation of shock fronts was reported \citep{grant:2018}. In this work, the highly inclined (70-80 degrees) magnetic field geometry in the outer boundary of the umbra, alongside the tangential velocity signatures, distinguished the waves as distinct from umbral flashes. Observed local temperature enhancements of 5$\%$ in the chromosphere were reported and thought to be evidence of dissipation of mode-converted Alfv\'en waves driven by upwardly propagating magneto-acoustic oscillations.  Alfvén shocks are predicted to form in regions with high negative Alfvén speed gradients \citep{hollweg:1982}.  In this case of the nearly horizontal outer umbral fields, the volume expansion of the magnetic fields and the fact that the density does not drop off steeply means a negative gradient in the Alfvén speed can exist. The study by \citet{grant:2018} motivated us to search for oscillatory signatures in highly inclined magnetic field geometries, even though our choice of feature being a polarity inversion line is a very different environment that that of the outer umbral field, we have an opportunity to characterize the transverse velocity oscillations.

The purpose of this paper is two-fold.  First, we explore the oscillations observed with HMI.  Since HMI observes the full solar disk for all of the magnetic activity present in Solar Cycle 24, and perhaps that of Solar Cycle 25, then any successful measurement of wave power as well as differentiation of photospheric wave modes from the use of HMI data will support other research conducted with higher spatial and temporal resolution instruments hosted by DKIST, SST, and other instruments around the world. Specifically, within this paper, we analyze the signals and oscillations within the active region NOAA 11158 on 16 Feb 2011 when the region was at South latitude 21$^{\circ}$ and West 35$^{\circ}$ longitude, a center-to-limb angle of 40.8$^{\circ}$. We use both the 45-second line-of-sight data and the 135-second vector data \citep{sun:2017}, a higher-cadence vector data set whose first release contains about 30 events and 290 hours of flaring regions.  Second, we observe Doppler velocity oscillations in AR11158 in the region of the polarity inversion line, where the magnetic fields are measured to be transverse to the observer, to detect Alfv\'enic motions.  We are not aware of other MHD wave studies conducted along a polarity inversion line. 

\section{Data \& Methods }

\subsection{Data}
We utilize data from the Helioseismic Magnetic Imager (HMI) aboard the Solar Dynamics Observatory (SDO).  HMI uses the 6173 Å line tuned to six wavelengths to produce 4096 $\times$ 4096 full disk images of Doppler velocity, line-of-sight magnetic field, line width and continuum intensity every 45 seconds with a pixel size of 0.5$^{\prime \prime}$ \citep{schou:2012, couvidat:2016}. This data can be found in JSOC with the names hmi.V$\_$45s, hmi.M$\_$45s, hmi.Lw$\_$45s, and hmi.Ic$\_$45s. The continuum intensity is a proxy determined from filtergram sampling away from line center. The line-widths vary due to thermal width changes and unresolved velocities within the same pixel.

 We analyzed a 512$\times$512 pixel region containing the AR11158 sunspot region on 2011 February 16 for eight hours from 09:00 - 17:00 UT. The line-of-sight data set with a cadence of one image every 45 seconds consists of 640 images of the line-of-sight magnetogram (B$_{los}$ or M), continuum intensity (I$_{c}$), line width (Lw), and Doppler velocity ($V$) images; the snapshots are shown in lower panels of Figure ~\ref{fig:vec}. The region is tracked by adjusting heliographic coordinates of map centers as a function of time using the mtrack routine which is available to the public through JSOC. The average center-to-limb angle of the region center is 40.8$^{\circ}$ at the beginning of the time interval. The data are stored as three-dimensional data cubes.
 The symbol, M, is used throughout this paper, the same as B$_{los}$ and indicates the magnetogram flux.  M was used to differentiate easily between the vector data quantity of field strength, B, from the inversion. While the line-width values do contain systematic errors \citep{cohen:2015,couvidat:2012}, changes in the thermal width and unresolved velocities within the same pixel increase the line width, and as such, it is valuable if we expect waves such as torsional Alfv\'en waves, and not compressional waves, to be present.
 
The Stokes I,Q,U,V are also recored by HMI at the six wavelength positions and by employing a Milne-Eddington inversion code, vector magnetic field maps are produced every 12 minutes \citep{centeno:2014, hoeksema:2014}.  These data can be found with the data name hmi.B$\_$720s with segments that include the field strength, inclination, azimuth, disambiguation, Doppler width and velocity. In addition to the 45 second line of sight data and the 720 second vector field data, there exists a recently produced vector data product with a shorter duration sampling, 135 seconds, for some time periods, with the data name hmi.B$\_$135s with all the segments produced by the inversion code \citep{sun:2017}.  The 135-second vector data is tracked for 8 hours consists of 213 images of field strength (B), inclination ($\gamma$), azimuth, Doppler width and velocity. 

\subsection{Neutral Line Determination}
We used a technique that measures the horizontal gradients of photospheric magnetic flux \citep{welsch:2008, schrijver:2007} in order to identify the polarity inversion line location. First, maps were created with values for regions where the magnitude of the magnetic field was over a certain threshold (20 G). The maps were dilated to allow for overlap of the positive and negative bitmaps, and this overlap  was taken to be the polarity inversion line. The locations were stored in an array, the "PIL bitmap" in which the locations of the neutral line has a value of 1 and elsewhere there was a value of 0.  We used the bitmaps, which were determined for every time step, to multiply the Doppler velocity images and other observables in order to create a neutral line data cube of each observable.  An example of the neutral line location is shown in Figure ~\ref{fig:pil}.  It is only the long portion between x of 200-300 and y of 280-310 that we consider the stable neutral line for this region.  Many of the pixels identified as PIL outside of the long portion are pixels located between plage that are evolving rapidly and as such, are not suitable to study as part of a time-series spanning several hours.    
 
\begin{figure}[!t]
\centering
\includegraphics[trim=0.8in 2.2in 1.65in 0.2in, clip=true,width=2.75in]{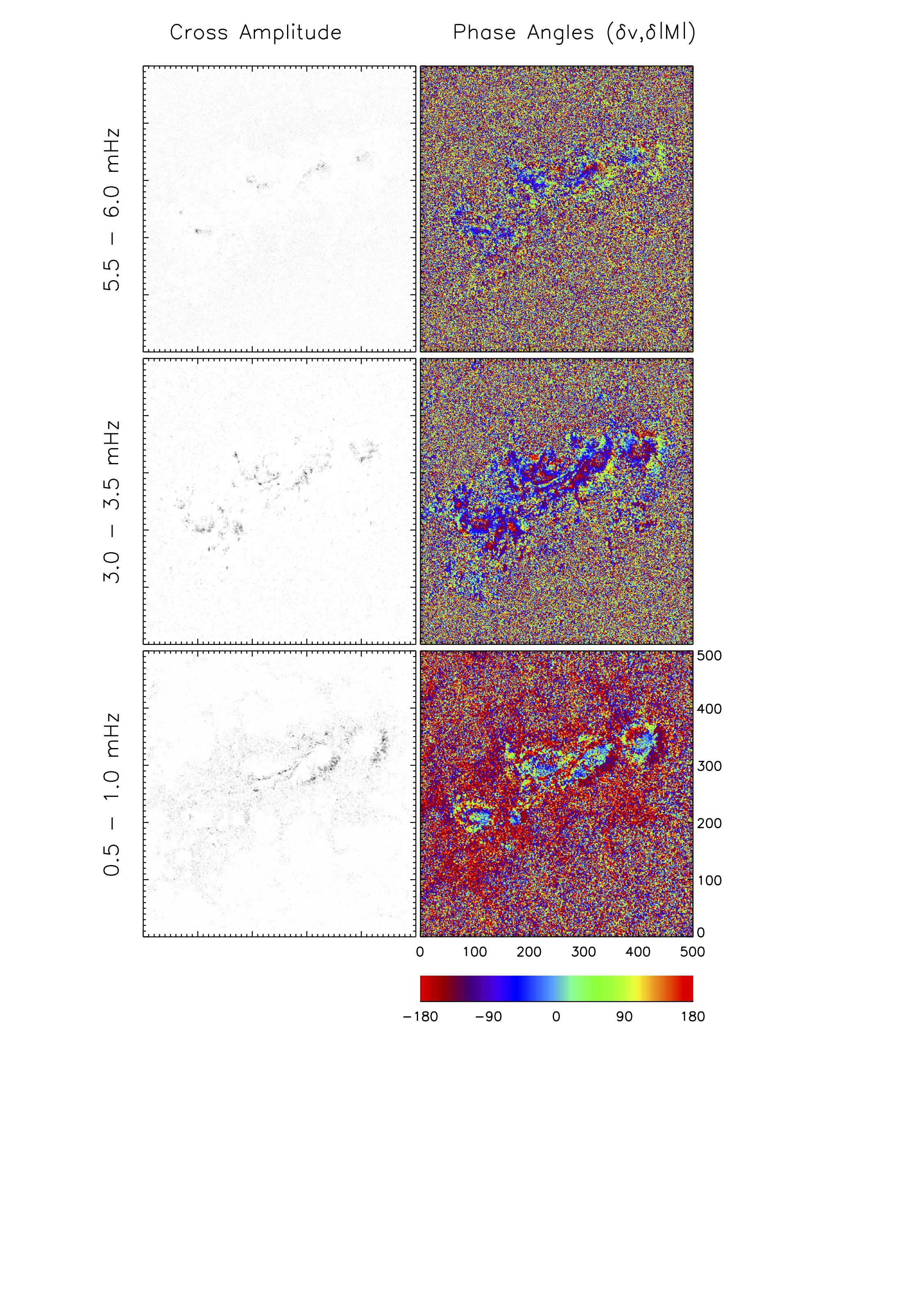}
\includegraphics[trim=0.8in 2.2in 1.65in 0.2in, clip=true,width=2.75in]{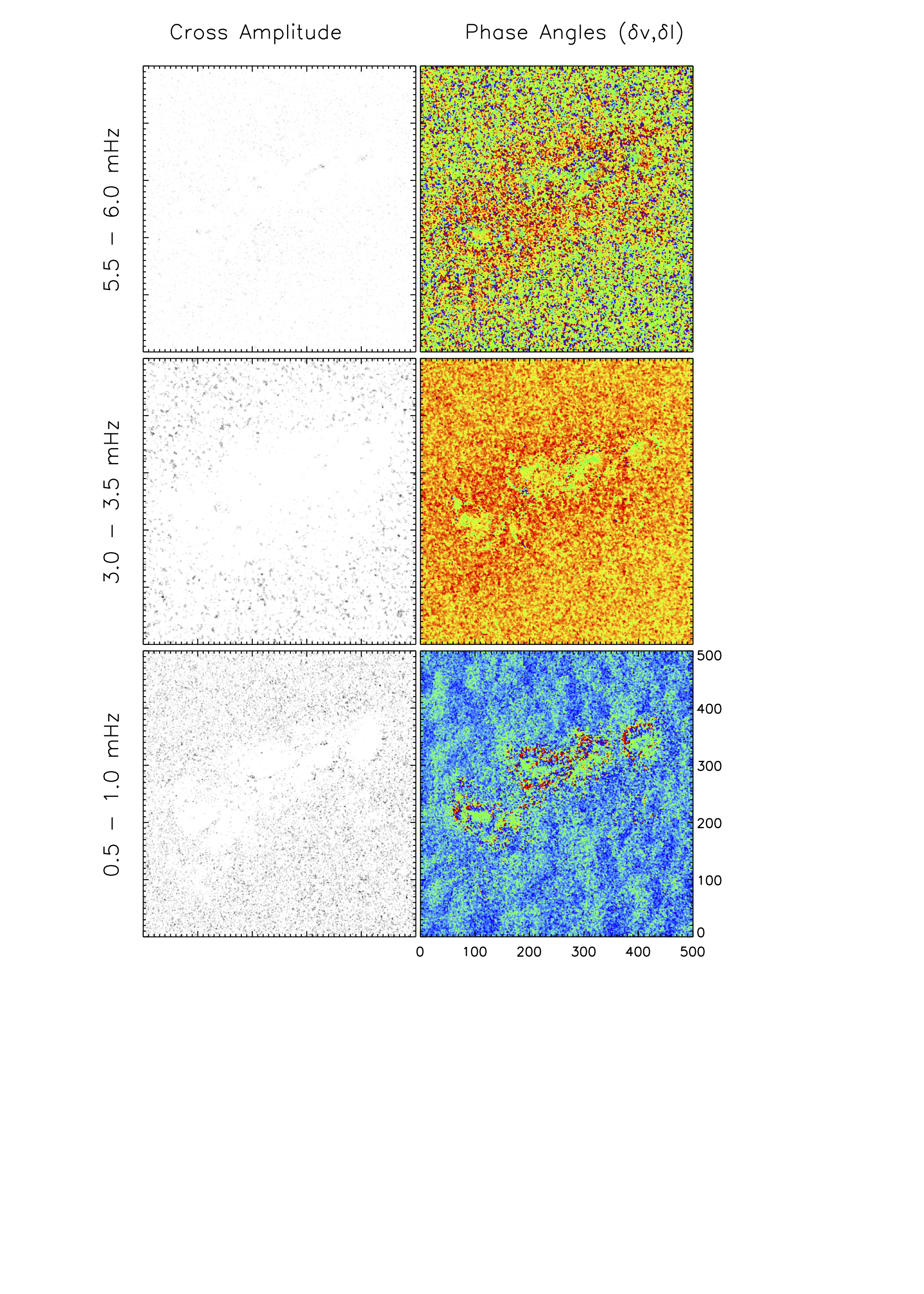}
\caption{The cross amplitude and phase values are shown for the Doppler velocity and B$_{los}$ signals in the left two columns. The cross amplitude and phase relationship are shown for the Doppler velocity and continuum intensity in the right two columns. Rows from top to bottom are the frequency ranges of 5.5-6.0, 3.0-3.5 and 0.5-1.0 mHz.}
\label{fig:spatial_phase}
\end{figure}

\subsection{Identification of Umbra, PIL, Plage and Quiet-Sun Subsets}
In order to characterize the oscillations in different features within the AR 11158 region, 160 pixels were selected for each of the features of umbra, PIL, plage and quiet-Sun.  Data was selected by isolating pixels in the region that matched the following criteria: PIL pixels needed to have field strengths between 600$-$1800 G, inclination between 75$-$105$^{\circ}$, and be located on the PIL in Figure ~\ref{fig:pil}; umbral pixels needed to have field over 2000 G and inclination less than 35$^{\circ}$; plage pixels needed to have field stregnths between 500$-$1000 G outside of the sunspot regions, and quiet-Sun pixels needed to have field under 150 G and inclination between 75$-$105$^{\circ}$. For a reminder of the physical conditions of these features, typical plasma $\beta$ values (the ratio of gas to magnetic pressure) for umbra, PIL, plage and quiet-Sun in the mid-photosphere are $\sim$ 0.7, 1$-$5 (similar to penumbra), 10$^{2}$, and 10$^{4}$, respectively \citep{gary:2001, mathew:2004, borrero:2011, cho:2017}. Average values and oscillation amplitudes of the features are shown in Table 1. The values for the 5-minute period are determined after average values have been removed and the data are filtered to remove all but the 2-4 mHz frequency interval.  The values for the 3-minute period use a 5-6 mHz frequency range for filtering. 

 \subsection{Fourier Analysis, Cross Spectra and Phase Determination}
 For the Fourier analysis, temporal variations from individual pixels are analyzed without averaging. The signals are de-trended using a Gaussian filter (30 minute width) before the average values are subtracted, see Figure~\ref{fig:osc} for sample oscillations.  
 The power spectra of the signals are computed. Three-dimensional data cubes are created where the temporal axis has been transformed into frequency.  The spatial distribution of power from the Fourier transform for the 512 $\times$ 512 region, averaged over select frequency ranges, is shown in Figure~\ref{fig:spatial_pow}, with context images of the line-of-sight data shown on the bottom row. 
 
 The time series is analyzed to determine the cross amplitude and phase angles for the various quantities, including the Doppler velocity and magnetic fluctuations, the Doppler velocity and continuum intensity, etc.
To compute the cross spectra, signals are interpolated onto a 10 second grid and shifted past each other in 10-second lag increments up  to  a $\pm$15.16  minute  time  interval.  The resulting cross covariance function is recorded. Restricting the  lag  interval  to  ensures  wave  train  coherence  time  is equivalent  to  applying  a  Bartlett  window.  The  Fourier transform  of  the  cross  covariance  function,  the  cross spectra, is computed at each position. Phases are determined from the arc-tangent of the imaginary over the real components of the cross spectra in the filtered frequency range. Taking the absolute value of the magnetic flux before cross-correlating signals ensure similar treatment for field variations, regardless of the magnetic polarity, thus eliminating a 180$^{\circ}$phase difference observed when not using the absolute measure of the magnetic flux. Cross amplitude and phase values are shown in Figure~\ref{fig:spatial_phase}. The Fourier power spectra for the 160 pixels that are representative of the umbral, PIL, plage and quiet-Sun features are shown in Figure~\ref{fig:summed_power} for the line-of-sight quantities of V, M, Ic and Lw.

Histograms of phase values are created for the 160 pixels for the selected  quiet, plage, PIL and umbral regions.  The bin size for the histograms is 20$^{\circ}$ and the resultant curves are shown in Figure ~\ref{fig:hist} for six relationships: (v,I), (v,$|$M$|$), (v,Lw), (I,$|$M$|$), (I,Lw) and (Lw,$|$M$|$).  Phases were not determined from the 135 second vector data since the spectra noise was higher.

\subsection{Wavelet Analysis}
Wavelet analysis has become a standard tool for identifying periodicities within time series \citep{torrence:1986}. The benefit of using Wavelet analysis as opposed to traditional Fourier methods is that wavelets allow for the determination of whether the oscillatory power varies over the duration of the observation. In Fourier analysis, the basis functions are localized in the frequency domain, whereas in wavelet analysis, they are localized in both the frequency and time domains. Information is then gathered about the amplitude of periodic signals and how this amplitude varies over the duration of the sampling.  For each pixel and observable (v, M or B$_{los}$, Ic and Lw), we perform a Morlet \citep{torrence:1986} wavelet transformation and use the 95$\%$ significance level to establish that the periods are real.  Figure~\ref{fig:wavelet1} shows representative results from quiet-Sun and umbral pixels while Figure~\ref{fig:wavelet2} shows results from PIL and plage.  Thin, vertical white lines are overplotted at the times when the region flared and the cone-of-influence, which denotes edge regions whose results are distorted, is also shown as a thin, white line visible in the lower corners of the plots.

\subsection{Time-Distance Slices}
In order to observe waves patterns in nearby pixels, we take a 30-pixel slice of data in the x-direction along the direction of the polarity inversion line, across an umbra and in quiet-Sun. We subtract the average values in each observable so that the oscillation amplitudes are apparent. We then stack these observations in time for two hours to create time-distance plots as seen in Figure ~\ref{fig:slits}. The background averages subtracted are shown in Figure~\ref{fig:slices}.  

\section{Results \& Discussion}

\subsection{Fourier Power \& Phase Relations}
Oscillations are observed in the HMI data at various locations with different amplitudes and phases, see Figure ~\ref{fig:osc} as an example of the velocity, intensity and magnetogram oscillations for a 70-minute period at a position along the PIL.  

The spatial distribution of power, shown in Figure ~\ref{fig:spatial_pow}, shows the evolution of the region and the locations of enhanced and suppressed wave power.
The 2nd to bottom row shows the 0.5-1.0 mHz power depicting evolutionary changes on the order of 20 minutes. While some features such as the sunspot umbra and polarity inversion line are fairly stable, plenty of evolution is seen such as the dispersion of plage (not easily tracked at a constant rate), radial outflow in penumbra, and general movement of small features in the surrounding area in the 0.5-1.0 mHz frequency range. The third row from the bottom depicts the 3.0-3.5 mHz band (5-minute period) and enhaced power is seen in the magnetogram and line width data in the active region while power is suppressed in the Doppler and intensity in the active region.  Quiet-Sun locations show enhanced 5-minute power in the velocity map. Regarding the 5.5-6.0 mHz range  (3-minute period), the small regions within the umbrae show enhanced magnetogram power while in the line-width data, there is 3-minute power seen in umbral, polarity inversion line and penumbral locations. The enhanced power in the 3-minute band in the Doppler data surrounding the active region is the acoustic halo. The grayscale range of the plots in Figure ~\ref{fig:spatial_pow} are varied between frequency ranges so features are apparent. 

In Figure~\ref{fig:spatial_phase}, two plots are shown for the amplitude of the cross spectra (left, grayscale panels) and resultant phases $\phi$ (right, colorful panels) for $\phi$(v,$|$M$|$) and $\phi$(v,I). While the phase values are most reliable in locations where significant power is present in both of the signals, all the phase values are shown in order to provide an idea of the structures in the data.  The supergranulation can be seen in the lower right $\phi$(v,I) panel on the order of 20 minutes in the green and blue patches, otherwise the 0.5-1.0 mHz phases are not easily interpreted.  In the 3.0-3.5 mHz range, the $\phi$(v,$|$M$|$) within the strong magnetic regions range from 0 to $-$180$^{\circ}$ whereas the $\phi$(v,I) values in these same locations are around 100$^{\circ}$. 
The quiet-Sun shows a random distribution of $\phi$(v,$|$M$|$) values and $\phi$(v,I) values clustered between 90-150$^{\circ}$. Signal is found in the umbrae for the 3-minute (5.5-6.0 mHz frequency) range (upper row) in the cross amplitudes of (v,$|$M$|$) and (v,I) with phases noisier but with similar values to the 5-minute oscillations. 

The average values and RMS oscillation amplitudes for the various features of umbra, PIL, plage and quiet-Sun are reported in Table 1.  The values were calculated as both single pixel values and averages of nearby pixels by applying a 2-dimensional Gaussian aperture with a 1.7 pixel FWHM.  The differences between the single pixel values and the spatially averaged ones were not significant, so only the single pixel values are reported.  The first four rows of Table 1 values are reported for the 5-minute period for all features, filtering signals to contain the 2-4 mHz frequencies only, prior to calculation.  The last row of Table 1 investigates the 3-minute oscillations seen in $\sim$20 umbral pixels, and filters the data for the 5.0-6.0 mHz frequency range prior to calculation.  We see suppression of p-mode amplitudes ($\delta$V) in the magnetic regions with the Doppler velocity amplitudes decreasing with increasing magnetic field strength such that velocity amplitudes in the quiet-Sun are highest and those in umbra are the lowest.  It is suspected that the acoustic waves are being converted into MHD waves here or that the magnetic tension is restricting the flux tubes being pushed around by the external gas motions. In Table 1, the I$_c$ oscillation amplitudes are similar in that the quiet-Sun shows the most variation, 1.5\% and the umbra shows the least, 0.8\%.   The RMS amplitudes of $\delta$B=35 Gauss and $\delta$M=12 Mx cm$^{-2}$ (excepting quiet-Sun) are very similar, even though the mean B and M values are quite different between features.  It may be that we are sampling the noise floor of the image, or sampling a cross talk, although the power spectra appear to show real, enhanced power in the 5-minute range.  The line width values are significantly higher in the PIL, with an average of 152 m\AA, than the values in the quiet-Sun, plage or umbral regions, 108, 122 and 128 m\AA, respectively.  The higher line width values are indicative of higher thermal widths and unresolved plasma motions. It is worth noting that the line width values are determined from filter positions closer to the core of the spectral line so they represent values somewhat higher in the atmosphere than those determined from the entire line profile, since the continuum is formed lowest in the atmosphere. The $\delta$Lw values are 1.4-1.8 m\AA~ excepting quiet-Sun, with the higher value being in the PIL.  Remember that RMS amplitudes are 0.7 the peak-to-peak amplitudes.  

\begin{figure}[!t]
    \centering
    \includegraphics[trim=0.00in 0.1in 0.2in 0.1in, clip=true, width=5.3in]{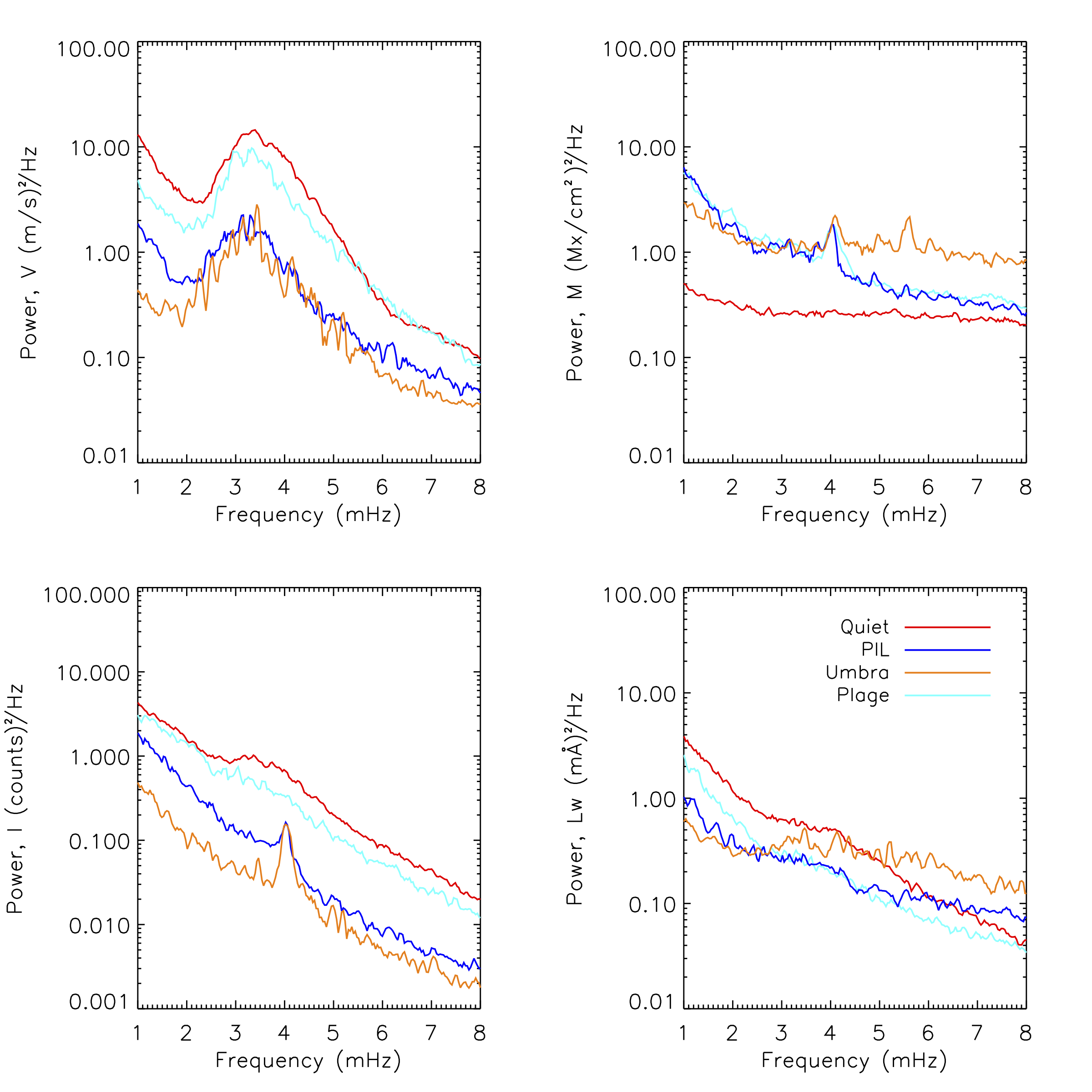}
    \caption{Average Fourier power, plotted on a log scale for the y-axis, for 160 pixels selected within the quiet-Sun, PIL, umbra, and  plage using 45 second cadence line-of-sight data. Clockwise from top left: Doppler power (divided by $10^{6}$), magnetogram power (divided by $10^{4}$), line width power (divided by $10^{3}$), and continuum power (divided by $10^{8}$). The 4 mHz peak is an artifact corresponding to the "pixel-crossing" time and is seen in many of the curves but no other artifacts are known. }
    \label{fig:summed_power}
\end{figure}

The average Fourier power computed for the features $-$ the umbra, PIL, plage and quiet-Sun $-$ using 45-second cadence data is shown in Figure~\ref{fig:summed_power}.  Significant 5-minute power is seen in all features in the velocity data, upper left panel, with amplitudes decreasing with increasing field strengths.  There is an artifact at 4 mHz which is the pixel-crossing time and is seen as a peak in magnetogram power in umbra, plage and PIL, in intensity power in the umbra and PIL data and in the umbral Lw data due to spatial gradients in the magnetic, intensity and line width data.  The tracking algorithm, mtrack, was employed to return a 0.03$^{\circ}$ pixel size and at $-$21$^{\circ}$ latitude, a pixel of 0.03$^{\circ}$ subtends 364 km.  At this latitude, the expected differential rotation rate is $\sim$1.57 km s$^{-1}$ but the sunspot rotates slightly slower at 1.46 km s$^{-1}$ and as such, it takes roughly 200 seconds for any feature to rotate into the next pixel, hence the 4 mHz power. The quiet-Sun magnetogram power (red line, upper right panel) represents pure noise. There is a peak at 5.6 mHz in the umbral magnetogram power that represents the 3-minute mode. The intensity power spectra (lower left panel in Figure~\ref{fig:summed_power}) shows 5-minute wave power in the quiet-Sun, some reduced 5-minute power in the plage, and the 4 mHz pixel crossing artifact in the PIL and umbra data. The line width power spectra (lower right panel) shows a bump of enhanced power in the 5-minute band for quiet-Sun and umbrae with perhaps some 3-minute wave power in the umbra, too, but it is fairly noisy.  There is a slight enhancement of 5-minute power for the plage and PIL data, too. 

\begin{figure}[!t]
    \centering
    \includegraphics[trim=0.07in 0.1in 0.2in 0.1in, clip=true, width=5.3in]{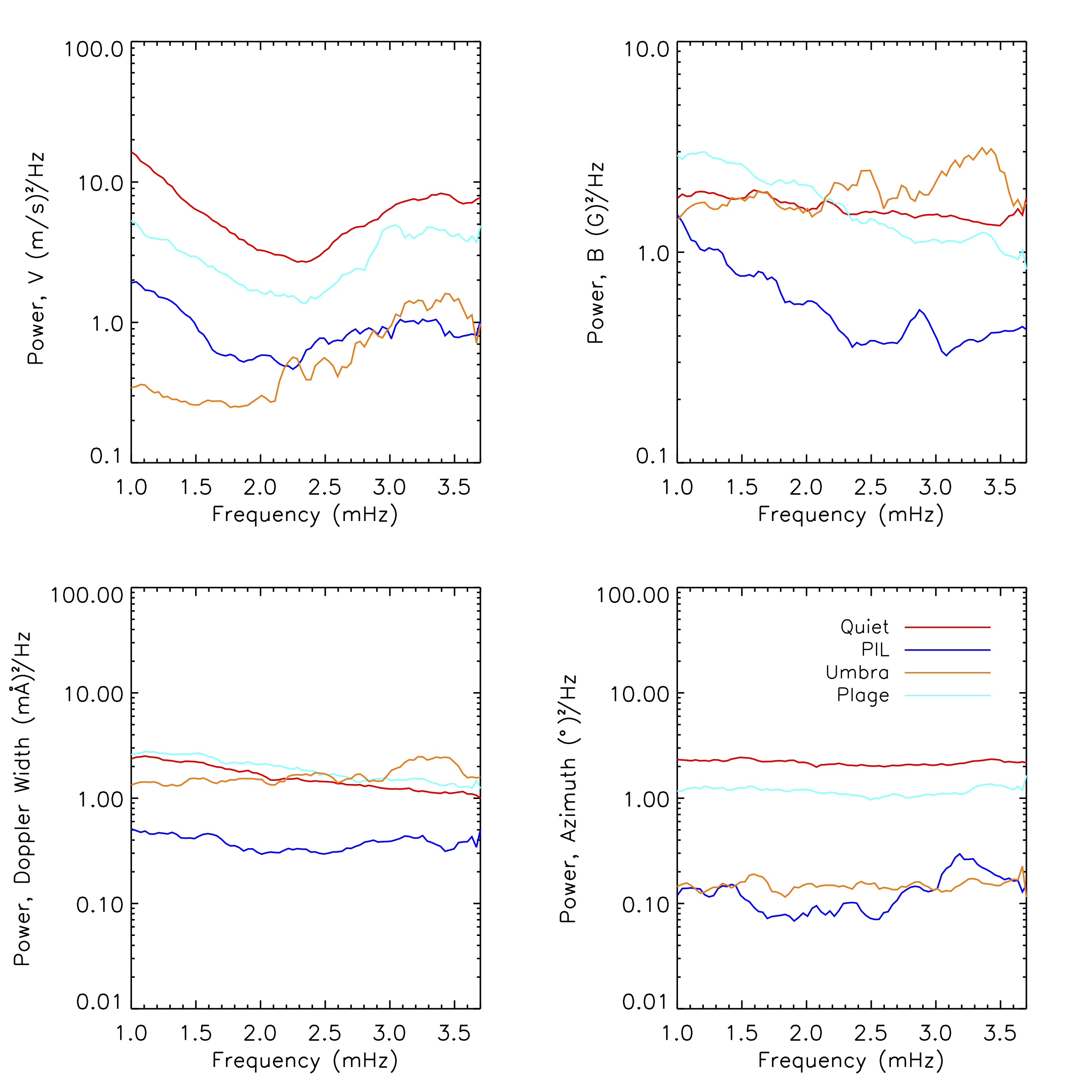}
    \caption{Average Fourier power, plotted on a log scale for the y-axis, for 160 pixels selected within the quiet-Sun, PIL, umbra, and  plage using 135 second cadence vector data. Clockwise from top left: Doppler (divided by $10^{6}$), magnetic field (divided by $10^{5}$), Doppler width (divided by $10^{3}$), and azimuth (divided by $10^{5}$). The inclination power spectra is not plotted as all curves appeared to be noise. The Nyquist frequency of this data is 3.7 mHz. The curves are smoothed by 2 points in frequency space as they are quite noisy.}
    \label{fig:summed_power_vec}
\end{figure}

The average Fourier power computed for the features $-$ the umbra, PIL, plage and quiet-Sun $-$ using 135-second cadence data from the vector field inversion is shown in Figure~\ref{fig:summed_power_vec}.  Significant 5- minute power is seen in all features in the velocity data, upper left panel, with amplitudes decreasing with increasing field strength features.  5-minute power is seen in the umbral field strength and Doppler width power spectra but very little other signal is seen. The azimuthal power spectra show no excess power, similar to the inclination power (not shown).  Phases are not calculated from the 135-second data as it is deemed too noisy for the phase values to be coherent.  

The mode, or most frequent value, of the phase for each feature and observable pair is shown in Table 2.  For all features, ${\phi}(v,I)=110-130^{\circ}$, indicating that the velocity signal leads the intensity signal for somewhat greater than a quarter $\frac{\pi}{2}$ of the $\sim$5-minute period, i.e. a lead of 120$^{\circ}$ for a 300 second period is 100 seconds. In contrast, the most frequent phase values of $\phi$(v,$|$M$|$)= $-$90$-$110$^{\circ}$ for PIL, umbra and plage indicate that the velocity signal lags the absolute value of the magnetic signal by a bit more than $\frac{\pi}{2}$. To interpret the phase values as wave modes, we use the Table 1 from \citet{moreels:2013} whose mathematical framework predicts that a $\phi(B_{los},v)=\pm\frac{\pi}{2}$ with a $\phi(v,I)=\pm\frac{\pi}{2}$ and $\phi(I,B_{los})=0,{\pi}$ are signatures of sausage waves that are slow standing or fast standing surface modes. The observed quantities, as used for determining these expected phase values, are line-of-sight velocity and magnetic flux. The HMI phase values shown in corresponding columns of Table 2 are consistent with these wave modes. 

\begin{figure}[!t]
\centering
\includegraphics[width=5.6in]{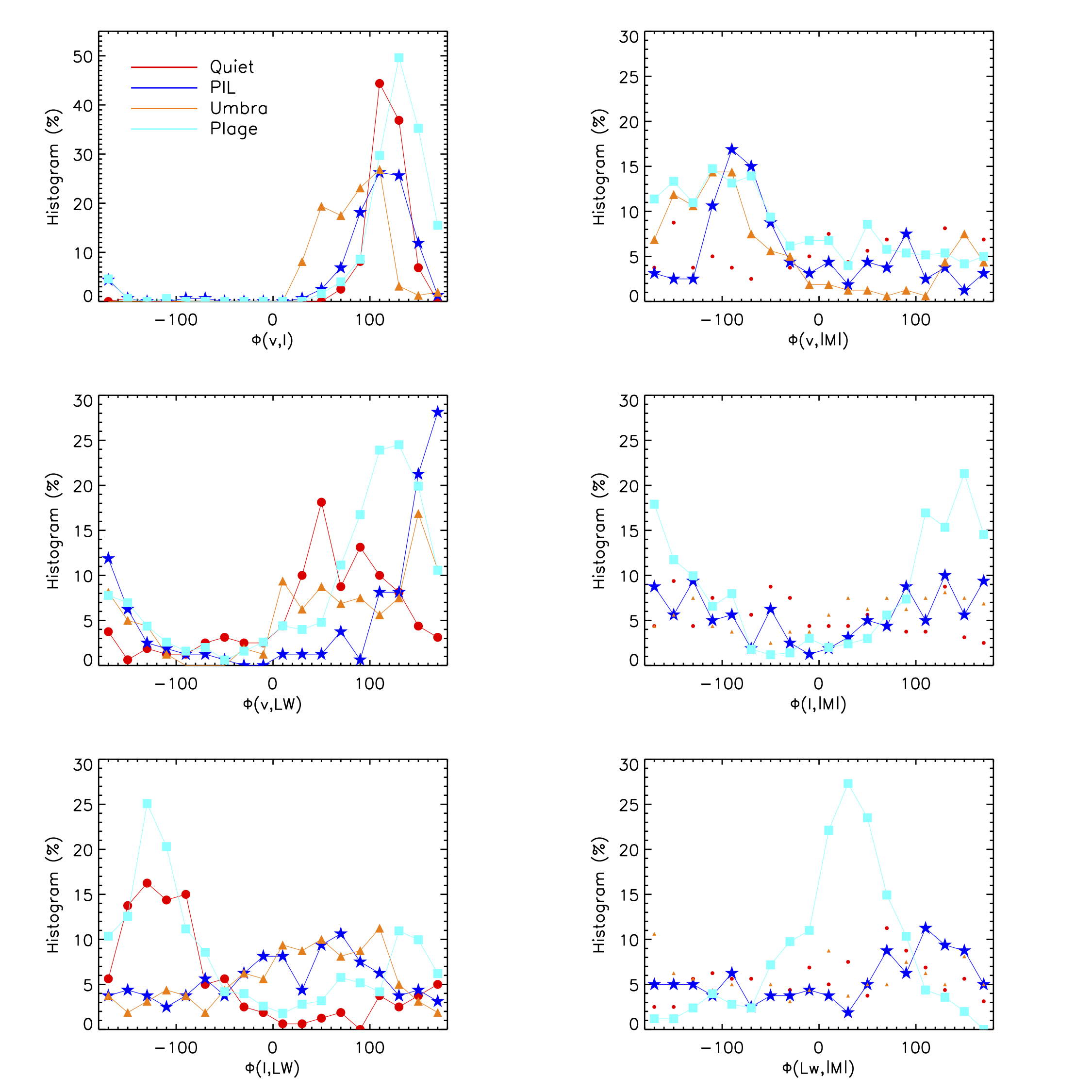}
\caption{Histograms with 20$^{\circ}$ bins are created for the phase values for oscillations of velocity and intensity, $\phi$(v,I), velocity and magnetic flux, $\phi$(v,$|$M$|$), velocity and line width, $\phi$(v,Lw), intensity and line width, $\phi$(I,Lw), and line width and magnetic flux, $\phi$(Lw,$|$M$|$) for pixels in the quiet-Sun, PIL, umbra and plage regions. Lines are not plotted for the quiet-Sun and plage data for panels in which the data are very noisy.  }
\label{fig:hist}
\end{figure}

Phases between the line-width and other observables vary more between the different features.  $\phi$(I,Lw) of umbra and PIL peak between 70-110$^{\circ}$ while plage and quiet-Sun peak at $-$130$^{\circ}$, see Figure ~\ref{fig:hist} lower left panel and corresponding column in Table 2. This indicates mechanisms that are 180$^{\circ}$ out of phase from each other in intensity and line width variations. $\phi$(Lw,$|$M$|$) values are 110$^{\circ}$ for PIL and 30$^{\circ}$ for plage, indicating modes that are $\frac{\pi}{2}$ out of phase from each other.  This could be geometric because the PIL is horizontal to the observer's line of sight with wave motions being perpendicular to the field direction whereas the plage is more radial and we are observing changes along the flux tube in the z direction.  

\begin{longtable}{|l|l|r|r|r|r|r|r|r|r|r|r|}
\caption{Average values and RMS amplitudes} \label{tab:osc} \\
\hline 
\multicolumn{1}{|c|}{\textbf{Period}} &
\multicolumn{1}{|c|}{\textbf{Feature}} &
\multicolumn{1}{c|}{\textbf{B}}  &
\multicolumn{1}{c|}{\textbf{\boldmath$\delta$B}} &
\multicolumn{1}{c|}{\textbf{\boldmath$\gamma$}}  &
\multicolumn{1}{c|}{\textbf{\boldmath$\delta\gamma$}}&
\multicolumn{1}{c|}{\textbf{M}}  &
\multicolumn{1}{c|}{\textbf{\boldmath$\delta$M}} &
\multicolumn{1}{c|}{\textbf{LW}}  &
\multicolumn{1}{c|}{\textbf{\boldmath$\delta$LW}} &
\multicolumn{1}{c|}{\textbf{\boldmath$\delta$v}}  & 
\multicolumn{1}{c|}{\textbf{\boldmath$\delta$I$_c$}} 
\\
\hline 
\rowcolor{LightCyan}
\textbf{5-minute} & Umbra & 2030 & 31& 23& 0.9 &
      1517 & 12 & 128 & 1.4 & 58 &0.8\% \\
 & PIL & 1213 & 38& 96& 0.8&
      $-$53 & 11 & 152 & 1.8 & 86 & 1.0\%\\
\rowcolor{LightCyan}
 & Plage & 602 & 38& 126$^{\dagger}$& 2.9&
      $-$388 & 13 & 122 & 1.4 & 126 &1.0\%\\
 & Quiet & \cellcolor{Gray}153 & \cellcolor{Gray}32& \cellcolor{Gray}90 & \cellcolor{Gray}2.6&
      \cellcolor{Gray}0.2 & \cellcolor{Gray}4.6 & 108  &2.4 & 222 & 1.5\%\\
      \hline
\rowcolor{LightCyan}
\textbf{3-minute} & Umbra & 2245 & -& 75 & - &
      1460 & 12 & 126 & 1.5 & 35 &0.3\% \\
\hline 
\end{longtable}
\footnotesize{Units are as follows: field strength, B (Gauss), inclination, $\gamma$ ($^{\circ}$), magnetic flux, M (Mx cm$^{-2}$), line width, LW (m\AA), velocity, v (m s$^{-1}$), intensity, I$_c$ (counts).  Gray-shaded cells are considered noise. Five (three) minute periods include 2-4 mHz (5-6 mHz) signals. $^{\dagger}$Inclinations of plage and small magnetic features, which have been shown to be nearly radial in the photosphere, are larger than expected in HMI data at high center-to-limb angles, which is due to Stokes Q and U having higher noise than Stokes V and the VFISV inversion weighting the Stokes profiles equally.} \\

\subsection{Wavelet Analysis}
Sample wavelet plots are shown in Figures~\ref{fig:wavelet1} and~\ref{fig:wavelet2}. In Figures~\ref{fig:wavelet1}, a quiet-Sun signal is shown in the top four panels and compared with an umbral signal shown in the bottom four panels. The quiet-Sun velocity shows 5-minute power peaking around 3.3 mHz, but ranging from 2.5-4.5 mHz, and the umbral panel shows significant power in a wider frequency band, from 2.5-5.8 mHz. Very little low-frequency power is seen in the umbral velocity. Less power is seen in the umbral intensity data than the quiet-Sun intensity, which is expected as the $p$-modes do modulate the intensity in the quiet-Sun. For the umbra, the magnetic wavelet panels show power in the 3-minute band (between 5-6 mHz) which is unexpected since the 3-minute umbral oscillation has only been previously observed in the Doppler data in the photosphere. The umbral line-width shows power at both the 5 and 3 minute band. Streaks of power that extend higher than 8 mHz are not understood.  Times of flares are indicated by thin, white vertical lines.  The plage and PIL wavelet plots in Figure~\ref{fig:wavelet2} look very similar to each other with plenty of low-frequency power in Ic, Lw and M and a band of significant power in the 5 minute range. 

\subsection{Time-Distance Analysis}
Time-distance data are shown in Figure~\ref{fig:slits} with a slit placed in the x-direction, sampling 30 pixels and stacked for 2 hours.  These are shown for a region across the PIL, umbra and quiet-Sun (columns left to right) and for Doppler, magnetogram and line-widths (rows top to bottom).  Background averages of the quantities are subtracted in order to show the fluctuations, but no filtering has been applied. The background averages are shown in Figure~\ref{fig:slices}. The Doppler velocities sampled across the PIL (top row, left panel) show strong variations that move at 2.7 km s$^{-1}$ in the plane-of-the-sky, seen in the top left panel as red or blue lines moving from the bottom right of the panel towards the upper left but terminating halfway. The magnetic oscillations shown in the PIL time distance data (middle row, left panel) begin their oscillations halfway across the slit, roughly where the Doppler oscillations weakened in the panel above.  The calculated velocities are 2-6 km/s for these bands. It is interesting to note that the line width variations (bottom row, left panel) are highest around the location of where the Doppler velocities across the PIL terminate and become magnetic oscillations.

The period of oscillations in magnetic flux across the PIL (left column, middle row panel in Figure 11) appear larger than those in the Doppler oscillations (left column, top row panel).  If the magnetic oscillations, which are likely a signature of Alfv\'en waves that have converted from acoustic oscillations along the right side of the PIL, then the differing periods of these two oscillation signals may contain important information on the nature of this conversion process. A relevant theoretical aspect to consider in this regard may be the resonant conversion of acoustic to Alfv\'enic waves at photospheric heights which has been discussed by \citet{zaqarashvili:2006, kuridze:2005} who both report findings that a sound wave is coupled to an Alfvén wave with double period and wavelength when the sound and Alfvén speeds are equal. Further investigation into the periods of magnetic and Doppler oscillations along the PIL are warranted. 

\begin{longtable}[!b]{|l|r|r|r|r|r|r|}
\caption{Summary of phase lags between observables} \label{tab:phase} \\
\hline \multicolumn{1}{|c|}{\textbf{Feature}} &
\multicolumn{1}{c|}{\textbf{\boldmath$\phi$(v,I)}}&
\multicolumn{1}{c|}{\textbf{(v,$|$M$|$)}}&
\multicolumn{1}{c|}{\textbf{(v,Lw)}}  &
\multicolumn{1}{c|}{\textbf{(I,$|$M$|$)}}&
\multicolumn{1}{c|}{\textbf{(I,Lw)}}  &
\multicolumn{1}{c|}{\textbf{(Lw,$|$M$|$)}} \\
\hline 
\rowcolor{LightCyan}
Umbra  & 110 & $-$110 & 150  &  $-$&  110   & $-$  \\ 
PIL    & 110 &  $-$90 & 170&   $-$ & 70   &110 \\
\rowcolor{LightCyan}
Plage  & 130 & $-$110 & 130& 150& $-$130  & 30\\ 
Quiet  & 110 & $-$    & 50 &  $-$&   $-$130  & $-$ \\
\hline 
\end{longtable}
\footnotesize{Histograms that do not show distinct peaks, i.e. do not having a most frequent phase value, have $-$ as a column entry. Phases are reported in units of degrees ($^{\circ}$) for oscillations in the 5-minute period (2$-$4 mHz range) such that a $-$90$^{\circ}$ phase value between (v,$|$M$|$) indicates that the velocity lags the absolute value of the magnetic signal by $\sim$75 seconds. The bin size is 20$^{\circ}$ and the results are reported with an inherent $\pm$10$^{\circ}$. \\

In the middle column of Figure~\ref{fig:slits}, the 3-minute umbral oscillations, with low amplitudes, can be easily seen in the Doppler data (top row, middle panel) and the 5-minute oscillations can be seen with high amplitudes in the quiet-Sun Doppler data (top row, right panel). Very little line width oscillations are seen in the umbra but strong line widths oscillations are seen in the quiet-Sun.  The PIL line width variations have finer structure as compared to the quiet-Sun line width variations, see lower row. The umbral and quiet-Sun time distance data show 3- and 5-minute oscillations but no similar behavior of acoustic to magnetic conversion, so we assume the geometry of the PIL with the horizontal, confined field lines provides an unusual environment for wave propagation. 

Of course, a concern in reporting amplitudes, power and phase values from a filtergraph type of instrument is that the time-varying signal is not measured per se, but is due to opacity or adiabatic fluctuations which then alter the height of formation or line profile characteristics in a way that cause periodic signals in the observables. Simulations need to be carried out with the 6173 \AA Fe-I line and the HMI data processing algorithms to quantify these effects, but this effort is outside the scope of this paper.

\begin{figure}[!t]
\includegraphics[width=4.5in]{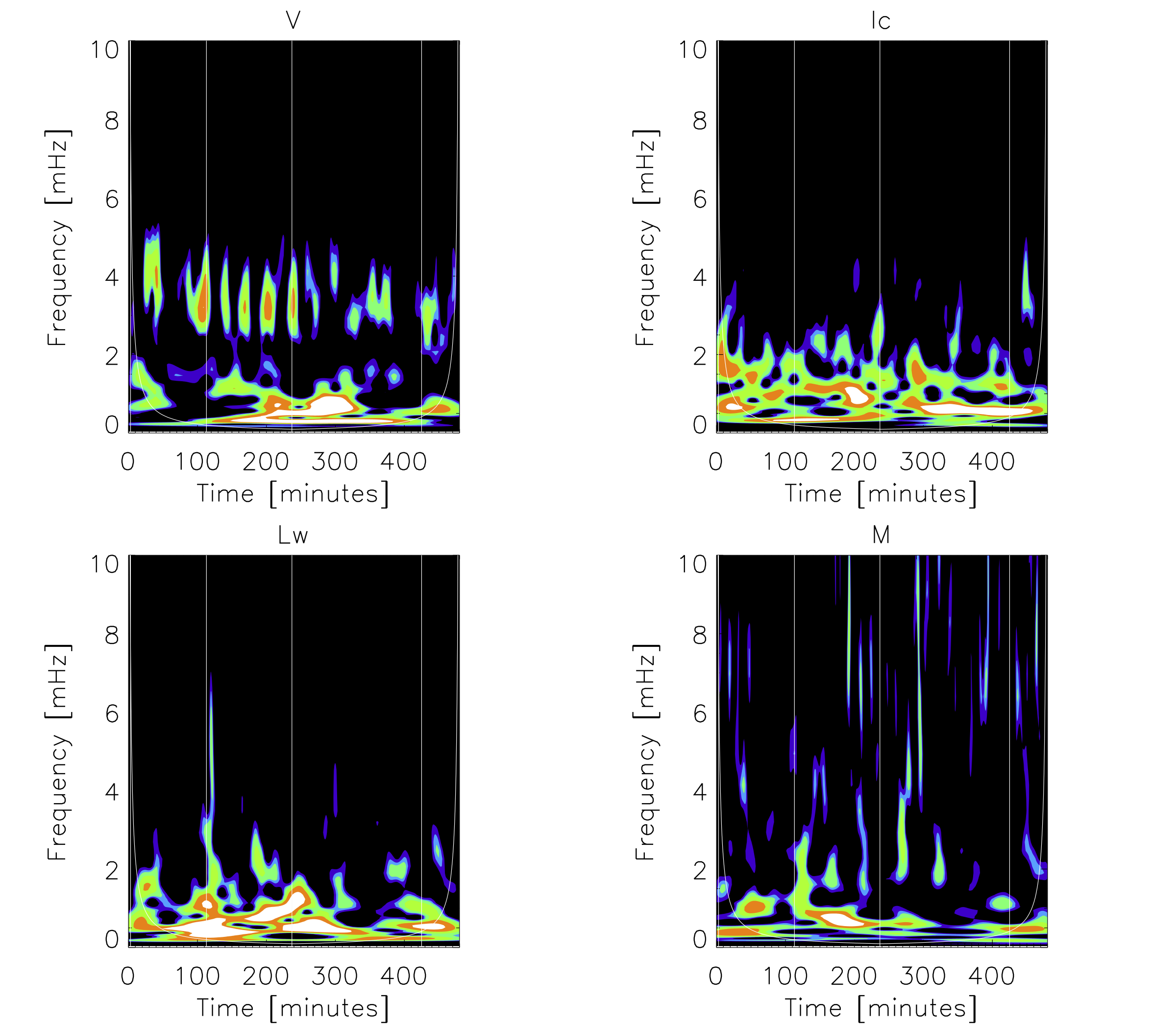}\\
\includegraphics[width=4.5in]{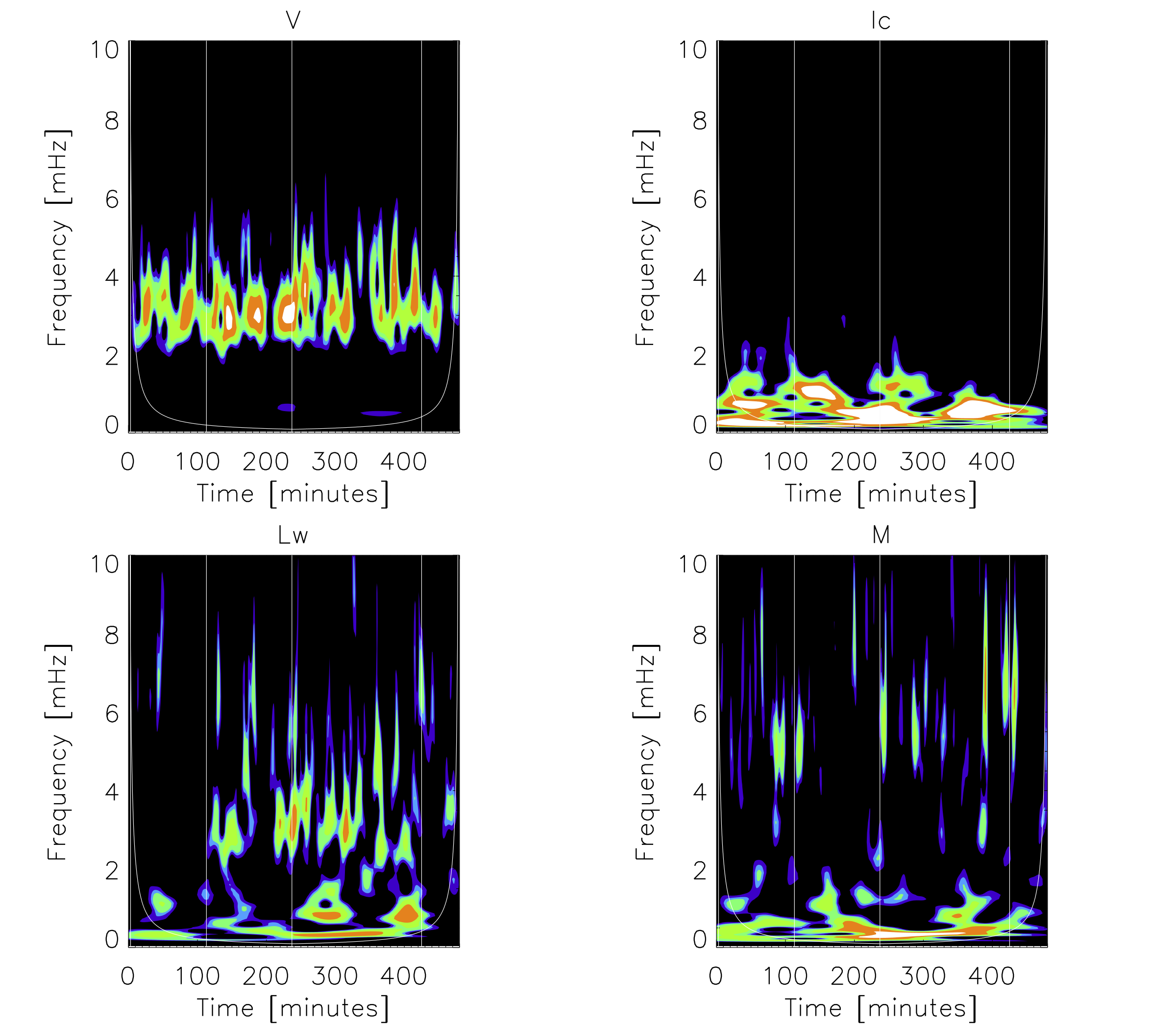}
\includegraphics[trim=0.05in 0.0in 6.in 0.00in, clip=true,width=0.75in]{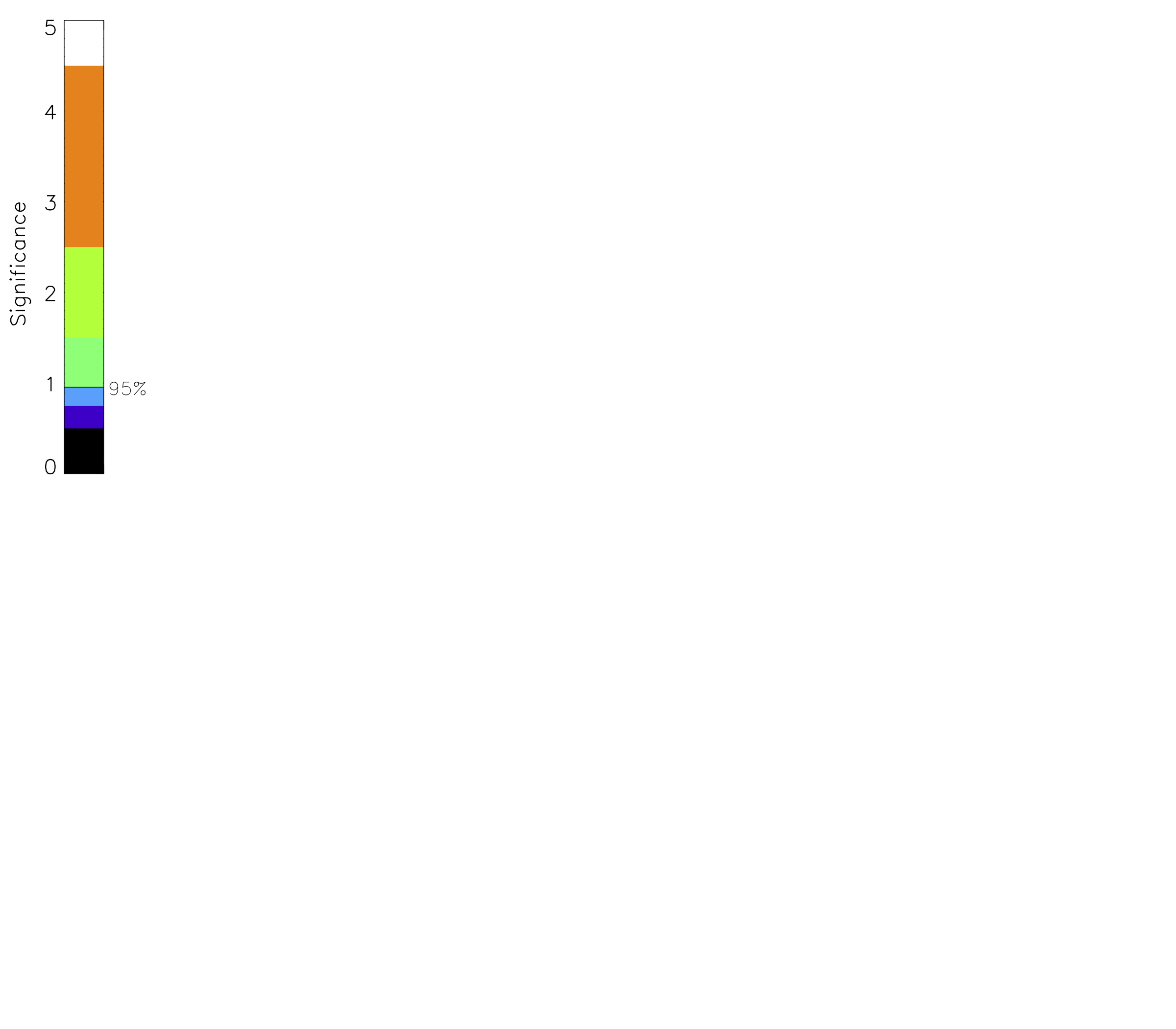}
\caption{Quiet-Sun (top four panels) and umbra (bottom four panels) wavelet power of Doppler velocity (V), intensity (Ic), line width (Lw) and magnetogram (M).  For the quiet-Sun signals, the field strength was lower than 100 Mx cm$^{-2}$.  For the umbral signal, the field strength was over 2000 Mx cm$^{-2}$ and the inclination was nearly radial at 166$^{\circ}$. The colorbar indicates the 7 contour level colors with the 95\% significant being between the blue and green shade. Thin, vertical white lines are overplotted at the times when the region flared and the cone-of-influence is also shown as a thin, white line visible in the lower corners of the plots.  }
\label{fig:wavelet1}
\end{figure}

\begin{figure}[!t]
\includegraphics[width=4.5in]{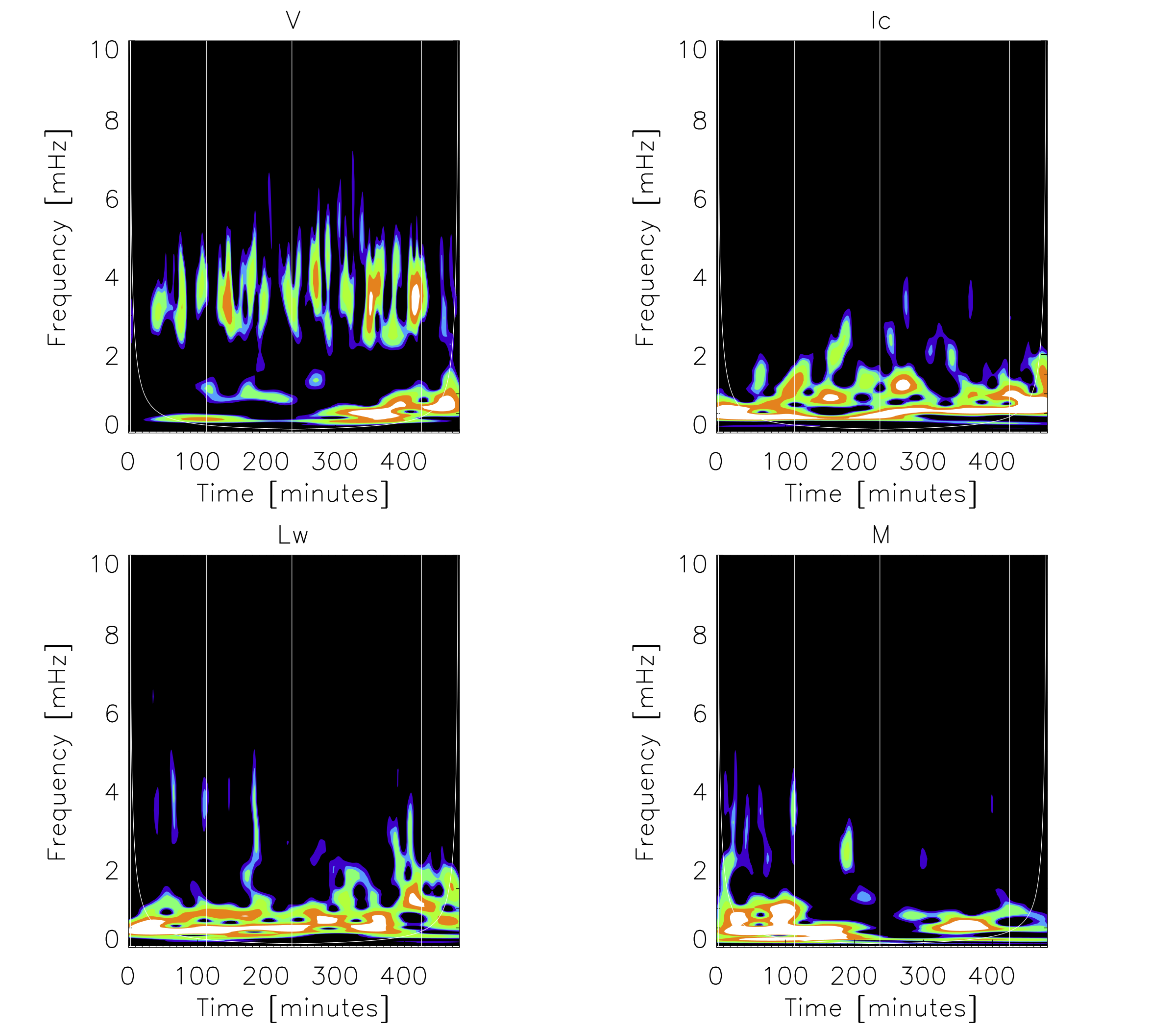}\\
\includegraphics[width=4.5in]{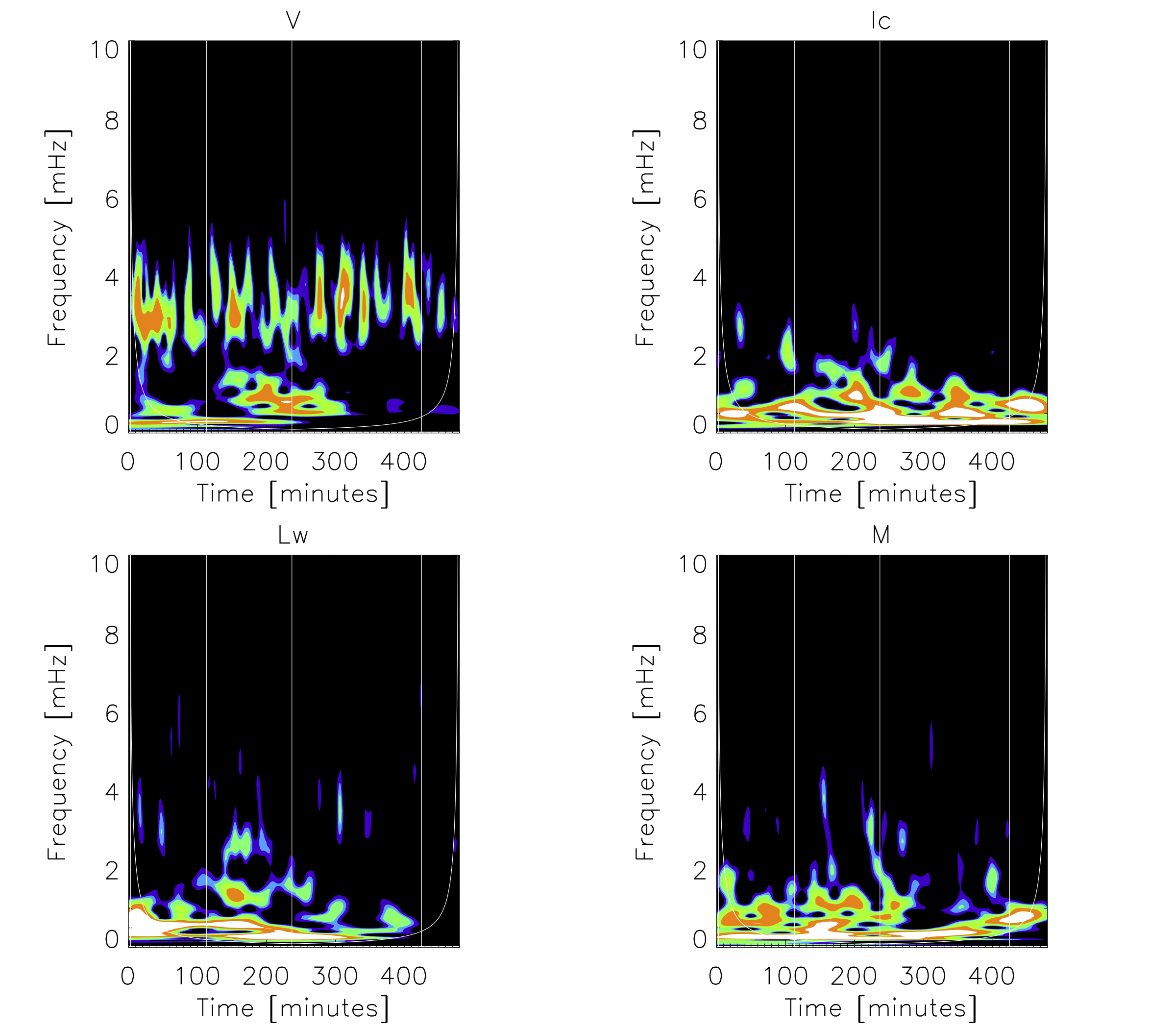}
\includegraphics[trim=0.05in 0.0in 6.in 0.00in, clip=true,width=0.75in]{images/fig9c.png}
\caption{Plage (top four panels) and polarity inversion line (bottom panels) wavelet power of Doppler velocity (V), intensity (Ic), line width (Lw) and magnetogram (M).  For the plage signals, the field strength was roughly 1000 Mx cm$^{-2}$ and the inclination 130$^{\circ}$.  For the PIL signal, the field strength was over 1600 Mx cm$^{-2}$ and the inclination was nearly horizontal at 97$^{\circ}$. The colorbar indicates the 7 contour level colors with the 95\% significant being between the blue and green. There is very little obvious difference between the PIL and plage signals.}
\label{fig:wavelet2}
\end{figure}

\begin{figure}[!ht]
\centering
\includegraphics[width=5.in]{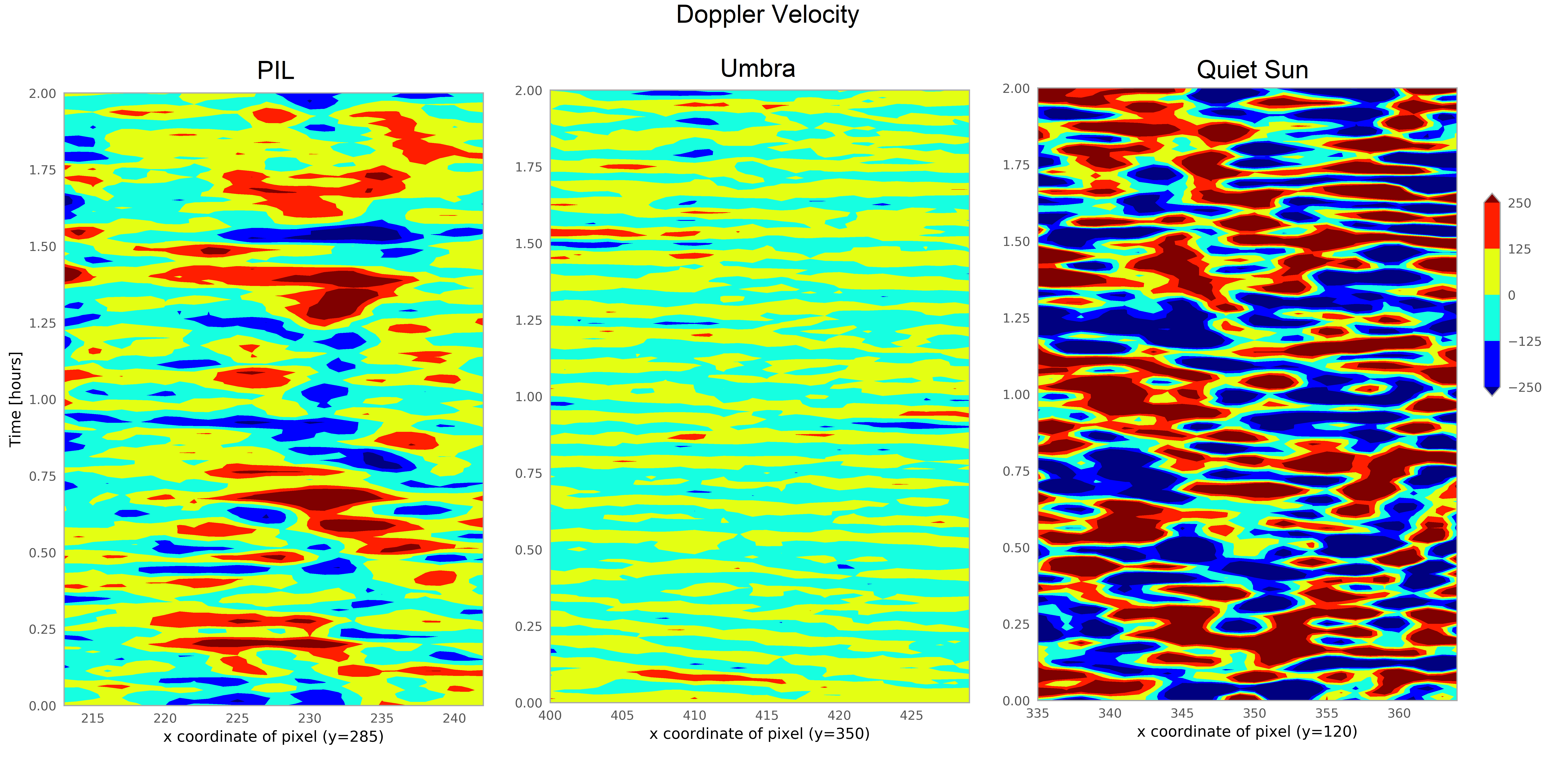}\\
\includegraphics[width=5.in]{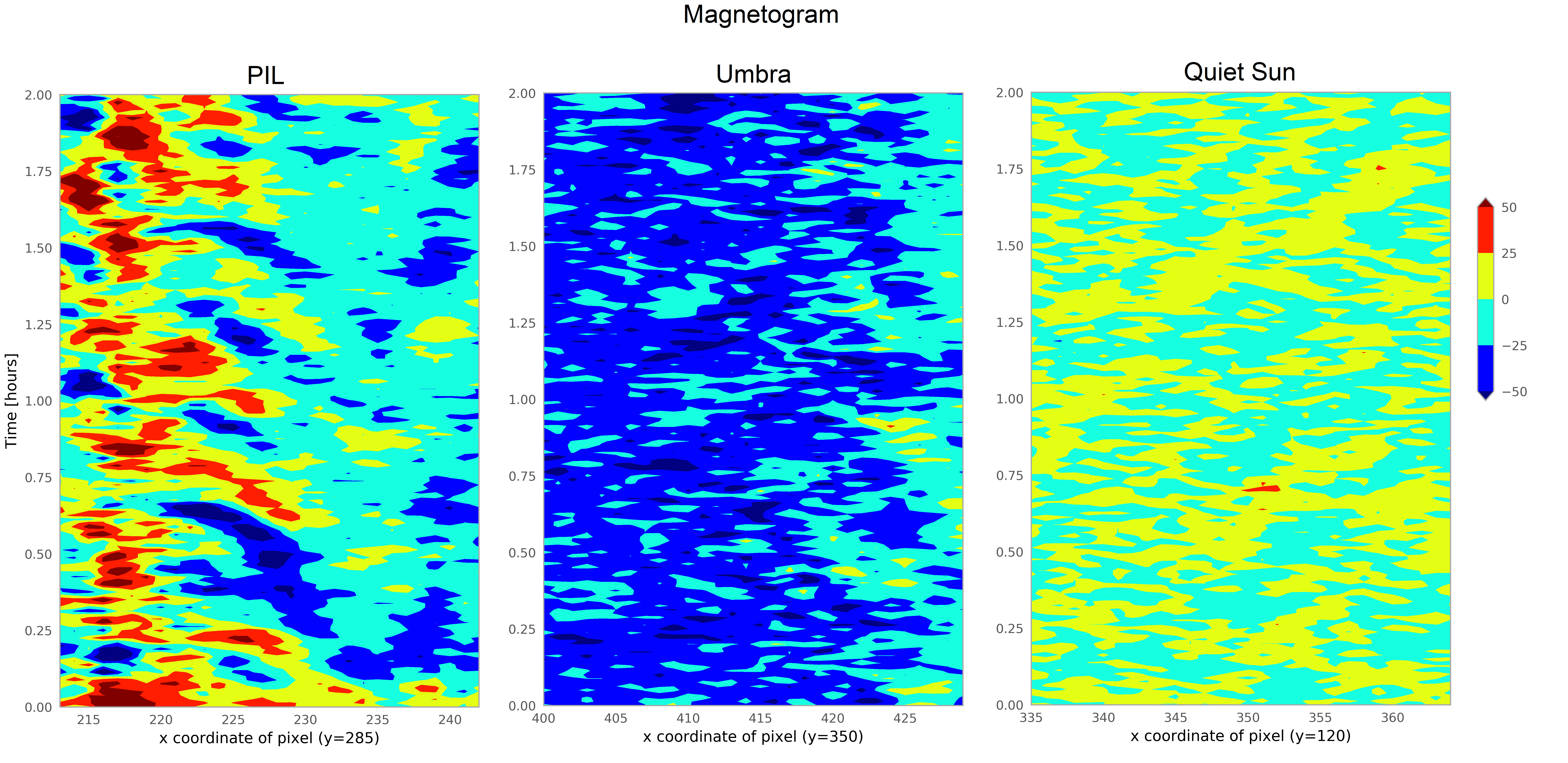}\\
\includegraphics[width=5.in]{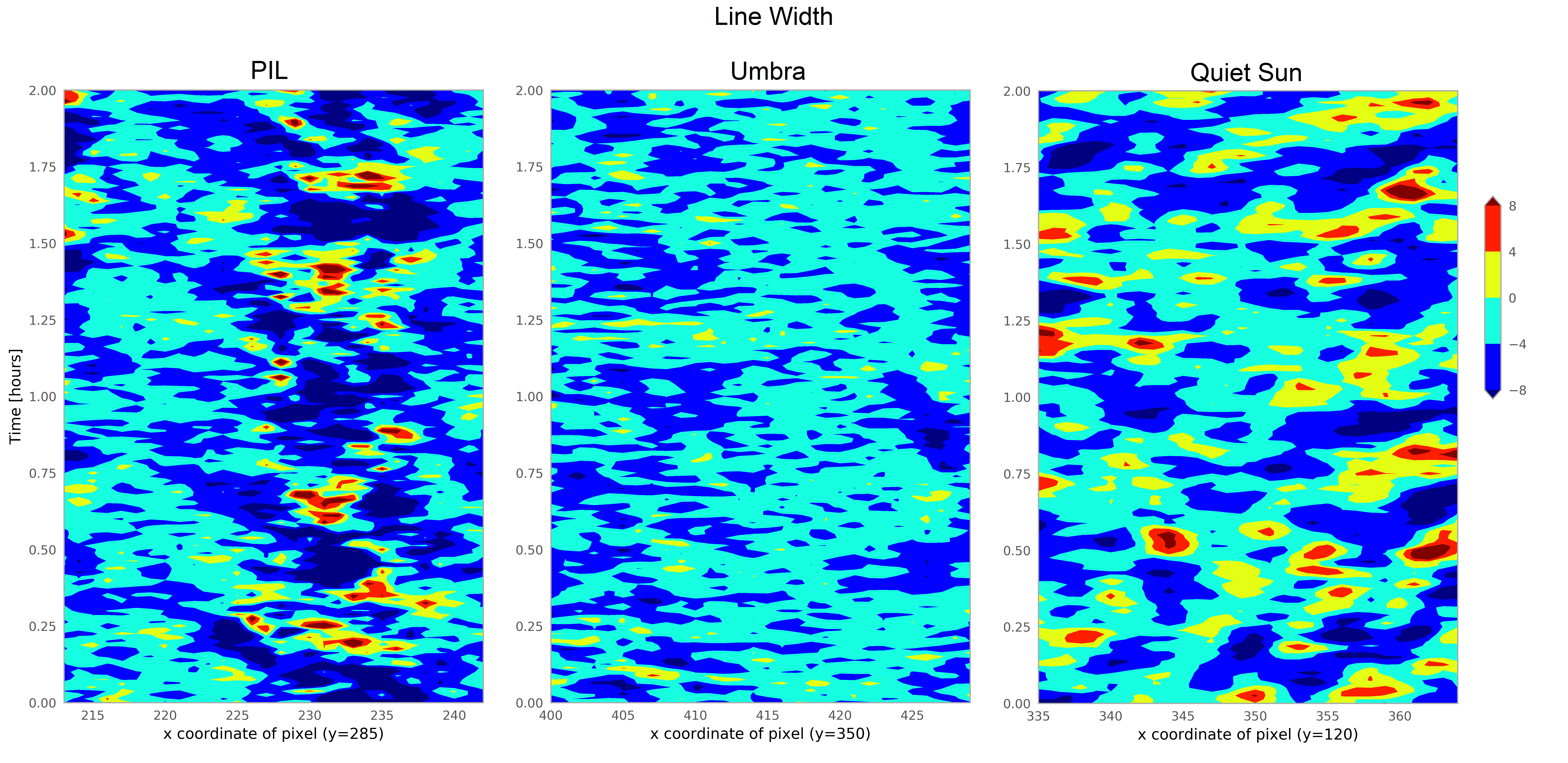}
\caption{Time-distance slices are shown for two hours for a slit placed horizontally along the polarity inversion line (left panels), a sunspot umbra (middle) and quiet-Sun (right).  The relative amplitude of the oscillations are shown in the Doppler velocity (top) with a range of $\pm$250 m s$^{-1}$, magnetic flux (middle) with a range of  $\pm$50 Mx cm$^{-2}$, and line width (bottom) with a range of  $\pm$8 m\AA.  Background averages have been subtracted in all cases but no filtering has been applied. }
\label{fig:slits}
\end{figure}

\begin{figure}[!ht]
\centering
\includegraphics[trim=0.0in 0.1in 4.3in 0.1in, clip=true, width=5.3in]{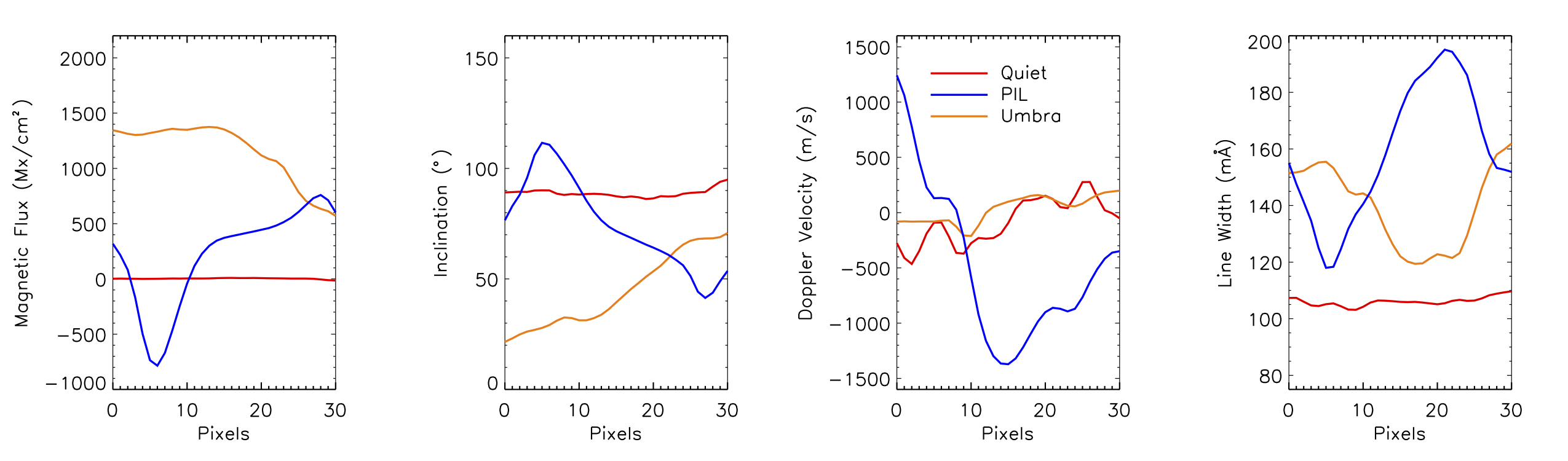}\\
\includegraphics[trim=4.25in 0.1in 0.0in 0.1in, clip=true, width=5.3in]{images/fig12.png}
\caption{The background averages of magnetic flux, inclination, Doppler velocity and line width for the two hours of time-distance data shown in Figure 11 for slices through quiet-Sun, PIL and umbral locations. The background averages are subtracted from the data in Figure 11 order to show the small-amplitude fluctuations. The inclination is included here to provide context for the geometry of the field in the PIL. When no magnetic field is present, as is the case for the quiet-Sun pixels, the inclination is near 90 degrees due to the higher noise in Q and U dominating the Stokes signals.}
\label{fig:slices}
\end{figure}

\section{Conclusions}

HMI provides a plentitude of data and the potential for its use is immense. In particular, the relatively high cadence of HMI for a full-disk imager allows one to see temporal variations in a large number of observables and thus makes it possible to study waves at frequencies relevant for the $p$-mode band and the 3-minute oscillations that permeate the photosphere and chromosphere and are suspected of driving at least some of coronal dynamics and energetics. This paper represents only a small fraction of what can be done with the data to investigate the presence of MHD waves in the photosphere.

In future efforts, we hope to conduct an analysis similar to that in this paper for an active region near disk center, or as an active region crosses the disk, to understand how the center-to-limb position affects the results.  Also, we hope to repeat this analysis using data corrected for scattered light.

In the vicinity surrounding AR 11158, and in the active region itself, we find the expected 5-minute power evident in the velocity and intensity data for quiet-Sun, see Figure~\ref{fig:summed_power}. The same is true for plage, although the power in the intensity is not as strong and we begin to see enhanced power in the magnetogram for plage. We also find significant oscillations in the 5-minute band for velocity signals from umbral and PIL pixels.  It is less evident, but power is enhanced in the 5-minute band for plage, PIL and umbral pixels in the magnetogram data and for all features in the line-width data.  The 135-second vector data confirms the umbral 5-minute oscillations in magnetic field and line width as the Doppler width and field strength of the vector data show a significant 5-minute peak, see Figure~\ref{fig:summed_power_vec}.

Surprisingly, for some umbral locations, a peak is seen in the magnetogram signal at the 3-minute period, around 5.6 mHz, in both the Fourier power (Figure~\ref{fig:summed_power}) and wavelet power spectra (Figure~\ref{fig:wavelet1}).Observation of 3-minute Doppler power has been reported in the photosphere in numerous studies, including IBIS observations of a  pore in the Fe I 617.3 nm line showed that 3-minute wave was already present at the photospheric height of formation of this line \citep{stangalini:2012}. The umbral 3-minute waves observed in the chromosphere were shown to be already present in the photosphere in a study by \citet{centeno:2009}. Numerical simulations \citep{felipe:2019} aided in the conclusion that propagation of waves in the 3-minute band directly from the photosphere can explain the observed chromospheric 3-minute oscillations.  So it is not shocking to see the 3-minute power, the surprising part is that it appears in the magnetogram data.

Phase values of HMI oscillations in ${\phi}(v,I)$, ${\phi}(v,|M|)$ and ${\phi}(I,|M|)$, shown in Table 2, observed in umbra, plage and PIL are in agreement with a mathematical framework of these waves being slow standing sausage modes.  This finding is consistent with numerous other studies which have interpreted 5-minute oscillations in magnetic structures in the photosphere as slow standing modes.

A time-distance diagram for a section across the PIL shows Doppler oscillations progressing Eastward at $\sim$2.7~km s$^{-1}$, with magnetic oscillation amplitudes increasing as the Doppler amplitudes damp, see Figure~\ref{fig:slits}, left column, top and middle rows.  
The magnetic disturbances then propagate at 2$-$6 km s$^{-1}$.
Enhanced line widths are found  at the locations where the waves change from being primarily acoustic to primarily magnetic. The umbral and quiet-Sun time distance data show 3- and 5-minute oscillations but no similar behavior of acoustic to magnetic conversion, so we assume the geometry of the PIL with the horizontal, confined field lines provides an unusual environment for wave propagation and a clear indication of the possible existence of Alfv\'en modes being generated and propagating in the PIL photosphere.

While the amplitudes of oscillations and phase relations in HMI data reported herein support the presence of MHD waves in and around the active regions, forward modeling of the spectral line dynamics in the presence of realistic MHD modes should be carried out prior to a final confirmation that the values reported in this paper are solar and not instrumental artifacts or crosstalk. 

Future collaboration within the Waves in the Lower Solar Atmosphere (WaLSA) group will help to isolate and characterize the photospheric wave properties here shown to be abundantly contained within the HMI data in order to support higher resolution studies from other instruments.  

\acknowledgements{This work was supported by NASA contract NAS5-02139 to Stanford University (HMI, PI P.H. Scherrer) and by a NASA Guest Investigator contract 80NSSC18L0668 (MHD Waves, PI J. Zhao). We wish to thank the Royal Society for hosting the Theo Murphy meeting at Chicheley Hall.  We also thank the Waves in the Lower Solar Atmosphere (WaLSA) group for their organization and support. Data are publicly available at \textit{jsoc.stanford.edu} and from the authors upon request.}

\clearpage
\bibliography{PILNorton}

\begin{thebibliography}{}
\expandafter\ifx\csname natexlab\endcsname\relax\def\natexlab#1{#1}\fi
\providecommand{\url}[1]{\href{#1}{#1}}

\bibitem[{{Alfv{\'e}n}(1947)}]{alfven:1947}
{Alfv{\'e}n}, H. 1947, Monthly Notices of the Royal Astronomical Society, 107,
  211

\bibitem[{{Bel} \& {Leroy}(1977)}]{bel:1977}
{Bel}, N., \& {Leroy}, B. 1977, \aap, 55, 239

\bibitem[{{Bellot Rubio} {et~al.}(2000){Bellot Rubio}, {Collados}, {Ruiz Cobo},
  \& {Rodr{\'\i}guez Hidalgo}}]{rubio:2000}
{Bellot Rubio}, L.~R., {Collados}, M., {Ruiz Cobo}, B., \& {Rodr{\'\i}guez
  Hidalgo}, I. 2000, \apj, 534, 989

\bibitem[{{Borrero} \& {Ichimoto}(2011)}]{borrero:2011}
{Borrero}, J.~M., \& {Ichimoto}, K. 2011, Living Reviews in Solar Physics, 8, 4

\bibitem[{{Braun} \& {Birch}(2008)}]{braun:2008}
{Braun}, D.~C., \& {Birch}, A.~C. 2008, Solar Physics, 251, 267

\bibitem[{{Braun} {et~al.}(1987){Braun}, {Duvall}, \& {Labonte}}]{braun:1987}
{Braun}, D.~C., {Duvall}, T.~L., J., \& {Labonte}, B.~J. 1987, The
  Astrophysical Journal Letters, 319, L27

\bibitem[{{Cally}(2005)}]{cally:2005}
{Cally}, P.~S. 2005, Monthly Notices of the Royal Astronomical Society, 358,
  353

\bibitem[{{Centeno} {et~al.}(2006){Centeno}, {Collados}, \& {Trujillo
  Bueno}}]{centeno:2006}
{Centeno}, R., {Collados}, M., \& {Trujillo Bueno}, J. 2006, \apj, 640, 1153

\bibitem[{{Centeno} {et~al.}(2009){Centeno}, {Collados}, \& {Trujillo
  Bueno}}]{centeno:2009}
---. 2009, \apj, 692, 1211

\bibitem[{Centeno {et~al.}(2014)Centeno, Schou, Hayashi, Norton, Hoeksema, Liu,
  Leka, \& Barnes}]{centeno:2014}
Centeno, R., Schou, J., Hayashi, K., {et~al.} 2014, \solphys, 289, 3531

\bibitem[{{Cho} {et~al.}(2017){Cho}, {Cho}, {Bong}, {Moon}, {Nakariakov},
  {Park}, {Baek}, {Choi}, {Kim}, \& {Lee}}]{cho:2017}
{Cho}, I.~H., {Cho}, K.~S., {Bong}, S.~C., {et~al.} 2017, \apjl, 837, L11

\bibitem[{{Cho} \& {Chae}(2020)}]{cho:2020}
{Cho}, K., \& {Chae}, J. 2020, \apjl, 892, L31

\bibitem[{{Cohen} {et~al.}(2015){Cohen}, {Criscuoli}, {Farris}, \&
  {Tritschler}}]{cohen:2015}
{Cohen}, D.~P., {Criscuoli}, S., {Farris}, L., \& {Tritschler}, A. 2015,
  \solphys, 290, 689

\bibitem[{{Couvidat} {et~al.}(2012){Couvidat}, {Schou}, {Shine}, {Bush},
  {Miles}, {Scherrer}, \& {Rairden}}]{couvidat:2012}
{Couvidat}, S., {Schou}, J., {Shine}, R.~A., {et~al.} 2012, Solar Physics, 275,
  285

\bibitem[{{Couvidat} {et~al.}(2016){Couvidat}, {Schou}, {Hoeksema}, {Bogart},
  {Bush}, {Duvall}, {Liu}, {Norton}, \& {Scherrer}}]{couvidat:2016}
{Couvidat}, S., {Schou}, J., {Hoeksema}, J.~T., {et~al.} 2016, Solar Physics,
  291, 1887

\bibitem[{{Felipe}(2012)}]{felipe:2012}
{Felipe}, T. 2012, The Astrophysical Journal, 758, 96

\bibitem[{{Felipe}(2019)}]{felipe:2019}
---. 2019, \aap, 627, A169

\bibitem[{{Felipe} \& {Sangeetha}(2020)}]{felipe:2020}
{Felipe}, T., \& {Sangeetha}, C.~R. 2020, arXiv e-prints, arXiv:2006.00526

\bibitem[{{Fujimura} \& {Tsuneta}(2009)}]{fujimura:2009}
{Fujimura}, D., \& {Tsuneta}, S. 2009, \apj, 702, 1443

\bibitem[{{Gary}(2001)}]{gary:2001}
{Gary}, G.~A. 2001, \solphys, 203, 71

\bibitem[{{Grant} {et~al.}(2015){Grant}, {Jess}, {Moreels}, {Morton},
  {Christian}, {Giagkiozis}, {Verth}, {Fedun}, {Keys}, {Van Doorsselaere}, \&
  {Erd{\'e}lyi}}]{grant:2015}
{Grant}, S.~D.~T., {Jess}, D.~B., {Moreels}, M.~G., {et~al.} 2015, The
  Astrophysical Journal, 806, 132

\bibitem[{{Grant} {et~al.}(2018){Grant}, {Jess}, {Zaqarashvili}, {Beck},
  {Socas-Navarro}, {Aschwanden}, {Keys}, {Christian}, {Houston}, \&
  {Hewitt}}]{grant:2018}
{Grant}, S. D.~T., {Jess}, D.~B., {Zaqarashvili}, T.~V., {et~al.} 2018, Nature
  Physics, 14, 480

\bibitem[{{Hoeksema} {et~al.}(2014){Hoeksema}, {Liu}, {Hayashi}, {Sun},
  {Schou}, {Couvidat}, {Norton}, {Bobra}, {Centeno}, {Leka}, {Barnes}, \&
  {Turmon}}]{hoeksema:2014}
{Hoeksema}, J.~T., {Liu}, Y., {Hayashi}, K., {et~al.} 2014, \solphys, 289, 3483

\bibitem[{{Hollweg} {et~al.}(1982){Hollweg}, {Jackson}, \&
  {Galloway}}]{hollweg:1982}
{Hollweg}, J.~V., {Jackson}, S., \& {Galloway}, D. 1982, \solphys, 75, 35

\bibitem[{{Jess} {et~al.}(2009){Jess}, {Mathioudakis}, {Erd{\'e}lyi},
  {Crockett}, {Keenan}, \& {Christian}}]{jess:2009}
{Jess}, D.~B., {Mathioudakis}, M., {Erd{\'e}lyi}, R., {et~al.} 2009, Science,
  323, 1582

\bibitem[{{Kanoh} {et~al.}(2016){Kanoh}, {Shimizu}, \& {Imada}}]{kanoh:2016}
{Kanoh}, R., {Shimizu}, T., \& {Imada}, S. 2016, \apj, 831, 24

\bibitem[{{Keys} {et~al.}(2018){Keys}, {Morton}, {Jess}, {Verth}, {Grant},
  {Mathioudakis}, {Mackay}, {Doyle}, {Christian}, {Keenan}, \&
  {Erd{\'e}lyi}}]{keys:2018}
{Keys}, P.~H., {Morton}, R.~J., {Jess}, D.~B., {et~al.} 2018, \apj, 857, 28

\bibitem[{{Khomenko} \& {Cally}(2012)}]{khomenko:2012}
{Khomenko}, E., \& {Cally}, P.~S. 2012, The Astrophysical Journal, 746, 68

\bibitem[{{Khomenko} {et~al.}(2008){Khomenko}, {Centeno}, {Collados}, \&
  {Trujillo Bueno}}]{khomenko:2008}
{Khomenko}, E., {Centeno}, R., {Collados}, M., \& {Trujillo Bueno}, J. 2008,
  \apjl, 676, L85

\bibitem[{{Khomenko} \& {Collados}(2006)}]{khomenko:2006}
{Khomenko}, E., \& {Collados}, M. 2006, The Astrophysical Journal, 653, 739

\bibitem[{Khomenko {et~al.}(2003)Khomenko, Collados, Solanki, Lagg, \&
  Bueno}]{khomenko:2003}
Khomenko, E.~V., Collados, M., Solanki, S.~K., Lagg, A., \& Bueno, J.~T. 2003,
  \aap, 408, 1115

\bibitem[{{Kuridze} {et~al.}(2005){Kuridze}, {Zaqarashvili}, \&
  {Roberts}}]{kuridze:2005}
{Kuridze}, D., {Zaqarashvili}, T.~V., \& {Roberts}, B. 2005, in ESA Special
  Publication, Vol.~11, The Dynamic Sun: Challenges for Theory and
  Observations, 89.1

\bibitem[{{Lites} {et~al.}(1998){Lites}, {Thomas}, {Bogdan}, \&
  {Cally}}]{lites:1998}
{Lites}, B.~W., {Thomas}, J.~H., {Bogdan}, T.~J., \& {Cally}, P.~S. 1998, \apj,
  497, 464

\bibitem[{{Mathew} {et~al.}(2004){Mathew}, {Solanki}, {Lagg}, {Collados},
  {Borrero}, \& {Berdyugina}}]{mathew:2004}
{Mathew}, S.~K., {Solanki}, S.~K., {Lagg}, A., {et~al.} 2004, \aap, 422, 693

\bibitem[{{Mathioudakis} {et~al.}(2013){Mathioudakis}, {Jess}, \&
  {Erd{\'e}lyi}}]{mathioudakis:2013}
{Mathioudakis}, M., {Jess}, D.~B., \& {Erd{\'e}lyi}, R. 2013, Space Science
  Review, 175, 1

\bibitem[{{Moreels} \& {Van Doorsselaere}(2013)}]{moreels:2013}
{Moreels}, M.~G., \& {Van Doorsselaere}, T. 2013, \aap, 551, A137

\bibitem[{{Norton} \& {Uitenbroek}(2002)}]{norton:2002}
{Norton}, A.~A., \& {Uitenbroek}, H. 2002, in ESA Special Publication, Vol.
  505, SOLMAG 2002. Proceedings of the Magnetic Coupling of the Solar
  Atmosphere Euroconference, ed. H.~{Sawaya-Lacoste}, 281--284

\bibitem[{{Norton} \& {Ulrich}(2000)}]{norton:2000}
{Norton}, A.~A., \& {Ulrich}, R.~K. 2000, \solphys, 192, 403

\bibitem[{{Norton} {et~al.}(2001){Norton}, {Ulrich}, \& {Liu}}]{norton:2001}
{Norton}, A.~A., {Ulrich}, R.~K., \& {Liu}, Y. 2001, \apj, 561, 435

\bibitem[{{Rajaguru} {et~al.}(2013){Rajaguru}, {Couvidat}, {Sun}, {Hayashi}, \&
  {Schunker}}]{rajaguru:2013}
{Rajaguru}, S.~P., {Couvidat}, S., {Sun}, X., {Hayashi}, K., \& {Schunker}, H.
  2013, \solphys, 287, 107

\bibitem[{{Rajaguru} {et~al.}(2019){Rajaguru}, {Sangeetha}, \&
  {Tripathi}}]{rajaguru:2019}
{Rajaguru}, S.~P., {Sangeetha}, C.~R., \& {Tripathi}, D. 2019, \apj, 871, 155

\bibitem[{{R\"uedi} {et~al.}(1998){R\"uedi}, {Solanki}, {Stenflo}, {Tarbell},
  \& {Scherrer}}]{Ruedi:1998}
{R\"uedi}, I., {Solanki}, S.~K., {Stenflo}, J.~O., {Tarbell}, T., \&
  {Scherrer}, P.~H. 1998, \aap, 335, L97

\bibitem[{Schou {et~al.}(2012)Schou, Borrero, Norton, Tomczyk, Elmore, \&
  Card}]{schou:2012}
Schou, J., Borrero, J.~M., Norton, A.~A., {et~al.} 2012, Solar Physics, 275,
  327

\bibitem[{{Schrijver}(2007)}]{schrijver:2007}
{Schrijver}, C.~J. 2007, The Astrophysical Journal Letters, 655, L117

\bibitem[{{Schunker} \& {Cally}(2006)}]{schunker:2006}
{Schunker}, H., \& {Cally}, P.~S. 2006, Monthly Notices of the Royal
  Astronomical Society, 372, 551

\bibitem[{{Settele} {et~al.}(2002){Settele}, {Carroll}, {Nickelt}, \&
  {Norton}}]{settele:2002}
{Settele}, A., {Carroll}, T.~A., {Nickelt}, I., \& {Norton}, A.~A. 2002, \aap,
  386, 1123

\bibitem[{{Stangalini} {et~al.}(2012){Stangalini}, {Giannattasio}, {Del Moro},
  \& {Berrilli}}]{stangalini:2012}
{Stangalini}, M., {Giannattasio}, F., {Del Moro}, D., \& {Berrilli}, F. 2012,
  \aap, 539, L4

\bibitem[{{Stangalini} {et~al.}(2018){Stangalini}, {Jafarzadeh}, {Ermolli},
  {Erd{\'e}lyi}, {Jess}, {Keys}, {Giorgi}, {Murabito}, {Berrilli}, \& {Del
  Moro}}]{stangalini:2018}
{Stangalini}, M., {Jafarzadeh}, S., {Ermolli}, I., {et~al.} 2018, \apj, 869,
  110

\bibitem[{{Sun} {et~al.}(2017){Sun}, {Hoeksema}, {Liu}, {Kazachenko}, \&
  {Chen}}]{sun:2017}
{Sun}, X., {Hoeksema}, J.~T., {Liu}, Y., {Kazachenko}, M., \& {Chen}, R. 2017,
  \apj, 839, 67

\bibitem[{{Torrence} \& {Compo}(1986)}]{torrence:1986}
{Torrence}, C., \& {Compo}, G. 1986, Bulletin of the American Meteorological
  Society, 79, 61

\bibitem[{{Ulrich}(1996)}]{ulrich:1996}
{Ulrich}, R.~K. 1996, \apj, 465, 436

\bibitem[{{Welsch} \& {Li}(2008)}]{welsch:2008}
{Welsch}, B.~T., \& {Li}, Y. 2008, in Astronomical Society of the Pacific
  Conference Series, Vol. 383, Subsurface and Atmospheric Influences on Solar
  Activity, ed. R.~{Howe}, R.~W. {Komm}, K.~S. {Balasubramaniam}, \& G.~J.~D.
  {Petrie}, 429

\bibitem[{{Werne} {et~al.}(2004){Werne}, {Birch}, \& {Julien}}]{werne:2004}
{Werne}, J., {Birch}, A., \& {Julien}, K. 2004, in ESA Special Publication,
  Vol. 559, SOHO 14 Helio- and Asteroseismology: Towards a Golden Future, ed.
  D.~{Danesy}, 172

\bibitem[{{Zaqarashvili} \& {Roberts}(2006)}]{zaqarashvili:2006}
{Zaqarashvili}, T.~V., \& {Roberts}, B. 2006, \aap, 452, 1053

\bibitem[{{Zhao} {et~al.}(2015){Zhao}, {Chen}, {Hartlep}, \&
  {Kosovichev}}]{zhao:2015}
{Zhao}, J., {Chen}, R., {Hartlep}, T., \& {Kosovichev}, A.~G. 2015, \apjl, 809,
  L15

\bibitem[{{Zhao} {et~al.}(2016){Zhao}, {Felipe}, {Chen}, \&
  {Khomenko}}]{zhao:2016}
{Zhao}, J., {Felipe}, T., {Chen}, R., \& {Khomenko}, E. 2016, \apjl, 830, L17

\end{thebibliography}
\end{document}